\documentclass[12pt]{article}
\usepackage[utf8]{inputenc}
\usepackage[T1]{fontenc}
\usepackage{helvet}
\usepackage{float}
\usepackage{authblk}
\usepackage{amsmath}
\usepackage{amssymb}
\usepackage{graphicx}
\usepackage{paralist}
\usepackage{mathtools}
\usepackage{fullpage}
\usepackage{xcolor}
\usepackage[font=footnotesize,width=0.8\textwidth]{caption} %
\usepackage{subcaption}
\usepackage{booktabs}
\usepackage{makecell}
\usepackage{enumitem}
\usepackage{changepage}
\usepackage{csquotes}
\usepackage{placeins}
\usepackage{hyperref}
\usepackage{adjustbox}
\usepackage{paralist}
\usepackage{xspace}
\usepackage{color}
\usepackage{balance}
\usepackage{colortbl}
\usepackage{longtable}
\usepackage{multirow}
\usepackage{comment}
\usepackage{microtype}
\usepackage{bbm}
\usepackage{appendix}

\newif\ifaddwatermark
\ifaddwatermark
    \usepackage{tikz}
    \usepackage{everypage}
    \AddEverypageHook{%
        \begin{tikzpicture}[remember picture,overlay]
            \node[anchor=north east, xshift=-2.5cm, yshift=-1cm] 
                at (current page.north east)
                {\textbf{\textcolor{red}{Draft — Do not share or cite without permission}}};
        \end{tikzpicture}%
    }
\fi

\usepackage[
backend=biber,
style=apa,
sorting=nyt,
citestyle=apa,
mincitenames=1,
maxcitenames=2,
uniquename=false,
uniquelist=false,
natbib=true]{biblatex}

\AtEveryBibitem{%
  \clearfield{issn} %
  \clearfield{isbn} %
  \clearfield{annotation} %
}

\DeclareSourcemap{
  \maps[datatype=bibtex]{
    \map{
      \pertype{misc}
      \step[fieldsource=howpublished, final]
      \step[fieldset=url, origfieldval, final]
      \step[fieldset=howpublished, null]
    }
  }
}

\DeclareCaptionFormat{custom}{#1#2#3}

\DeclareCaptionLabelSeparator{custom}{ | }

\DeclareCaptionFormat{supplementary}{#1#2#3}

\newif\ifSupplementaryStandalone
\SupplementaryStandalonefalse    %

\ifSupplementaryStandalone 
\usepackage{xr}
\makeatletter
\newcommand*{\addFileDependency}[1]{%
  \typeout{(#1)}
  \@addtofilelist{#1}
  \IfFileExists{#1}{}{\typeout{No file #1.}}
}
\makeatother
\newcommand*{\myexternaldocument}[1]{%
    \externaldocument{#1}%
    \addFileDependency{#1.tex}%
    \addFileDependency{#1.aux}%
}
\myexternaldocument{supplementary_standalone}
\fi

\hypersetup{
    colorlinks,
    linkcolor={red!50!black},
    citecolor={blue!50!black},
    urlcolor={blue!80!black}
}

\definecolor{lightblue}{HTML}{C5DCFA}

\makeatletter 
\renewcommand\AB@affilsepx{, \protect\Affilfont}
\makeatother

\newcommand{\para}[1]{{\vspace{7pt} \bf \noindent #1 \hspace{10pt}}}

\newenvironment{packed_enumerate}{
\begin{enumerate}
  \setlength{\itemsep}{2pt}
  \setlength{\parskip}{0pt}
  \setlength{\parsep}{0pt}
  \setlength{\topsep}{2pt}
}{\end{enumerate}}

\renewcommand{\mkbegdispquote}[2]{\itshape} %



\renewcommand{\arraystretch}{1.5}

\newcolumntype{L}[1]{>{\raggedright\arraybackslash}p{#1}}

\makeatletter
\renewcommand{\maketitle}{
    \begin{flushleft}
        \hspace{-0.5em}
        {\Large \@title \par}
        \vskip 1.5em
        \hspace{-1em}
        {\normalsize 
         \renewcommand{\arraystretch}{0}
         \begin{tabular}[t]{l} 
            \@author, \footnotesize *Equal contribution authors
         \end{tabular}
         \par}
        \vskip 0.5em
        {\@date}
    \end{flushleft}
    \par
    \vskip 1em
}
\makeatother

\renewenvironment{abstract}
 {\par\noindent\ignorespaces}
 {\par\medskip}

\renewcommand\Affilfont{\fontsize{10}{10}\selectfont}

\title{Data Voids and Warning Banners on Google Search}
\author[1]{Ronald E. Robertson$^*$}
\author[2]{Evan M. Williams$^*$}
\author[2]{Kathleen M. Carley}
\author[1]{David Thiel}
\affil[1]{Stanford University}
\affil[2]{Carnegie Mellon University}
\date{}

\begin{document}

\phantomsection
\addcontentsline{toc}{subsection}{Abstract}

\maketitle
\begin{abstract}
The content moderation systems used by social media sites are a topic of widespread interest and research, but less is known about the use of similar systems by web search engines. For example, Google Search attempts to help its users navigate three distinct types of data voids---when the available search results are deemed low-quality, low-relevance, or rapidly-changing---by placing one of three corresponding warning banners at the top of those search results. Here we collected 1.4M unique search queries shared on social media to surface Google's warning banners, examine when and why those banners were applied, and train deep learning models to identify data voids beyond Google's classifications. Across three data collection waves (Oct 2023, Mar 2024, Sept 2024), we found that Google returned a warning banner for about 1\% of our search queries, with substantial churn in the set of queries that received a banner across waves. The low-quality banners, which warn users that their results ``may not have reliable information on this topic,'' were especially rare, and their presence was associated with low-quality domains in the search results and conspiracy-related keywords in the search query. Low-quality banner presence was also inconsistent over short time spans, even when returning highly similar search results. In August 2024, low-quality banners stopped appearing on the SERPs we collected, but average search result quality remained largely unchanged, suggesting they may have been discontinued by Google. %
Using our deep learning models to analyze both queries and search results in context, we identify 29 to 58 times more low-quality data voids than there were low-quality banners, and find a similar number after the banners had disappeared. Our findings point to the need for greater transparency on search engines' content moderation practices, especially around important events like elections.
\end{abstract}

{\noindent \footnotesize \\ Correspondence and requests should be sent to rer@acm.org and evanmwilliams@cmu.edu} \\

\section{Introduction}

The content moderation practices of large online platforms, and how they're applied to information on important topics like health and elections, are a topic of widespread interest to researchers, the public, and policymakers around the world. 
One practice that has been widely debated is the use of ``warning labels'' to alert and inform users about the accuracy, context, or quality of a specific piece of content.
For example, social media sites have placed warning labels on posts classified as inaccurate by fact-checkers, and recent research suggests that such labels can be effective for reducing the belief and spread of misinformation~\citep{martel2023misinformation}.
However, the policies detailing when a platform will use a warning label are often opaque or absent~\citep{krishnan2021research}, can rapidly change, and research on how such labels are applied in practice has often been limited due to the challenges of surfacing examples to study~\citep{bradshaw2023investigation}.

\begin{figure}[t!]
\centering
\includegraphics[width=\textwidth, trim={0.05cm 0.05cm 0.05cm 0cm}, clip]{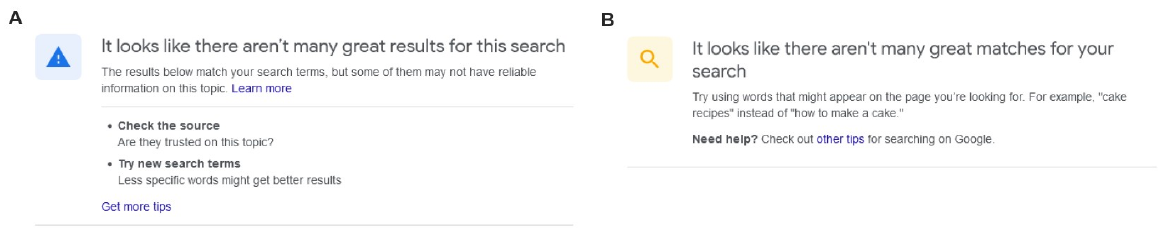}
\caption{Examples of warning banners for low-quality (A) and low-relevance (B) data voids on Google Search. Google displays the low-quality banner (A) at the top of its results when their ``systems don't have high confidence in the overall quality of the results available for the search''~\citep{nayak2022new}, and the low-relevance banner (B) ``when Google hasn't been able to find anything that matches your search particularly well''~\citep{tucker2020getting}. Google also has a rapidly-changing banner, but these are rare due to their time-sensitive nature and not a focus of this study (see Appendix~\ref{sec:appendix-banners-rapidly}, Figure~\ref{fig:freshness_banner}).}
\label{fig:banner_ex}
\end{figure}

Research in this vein has often focused on social media, but search engines also play a central role in online information seeking and employ similar content moderation practices~\citep{urman2024user}.
On Google Search, a close equivalent to the warning labels used on social media are the warning banners that Google places at the top of its search results to address \textit{data voids}: when the search results available for a given query are scarce, unstable, or dominated by unreliable or irrelevant websites~\citep{golebiewski2019data}. 
Such data voids are concerning because research suggests that search engines are widely trusted, but can increase belief in misinformation~\citep{aslett2024online}, influence political preferences~\citep{epstein2015search}, and be exploited as an indirect intermediary for guiding people into data voids~(\cite{golebiewski2019data}; \cite{tripodi2022propagandists}).
Google uses three distinct warning banners (Figure~\ref{fig:banner_ex}) to combat low-quality~\citep{nayak2022new}, low-relevance~\citep{tucker2020getting}, and rapidly-changing data voids~\citep{sullivan2021new}.
Similar to warning labels in social media, the use of warning banners in web search appears to generally help users evaluate low-quality or biased search results (\cite{epstein2017suppressing}; \cite{ludolph2016manipulating}), though warnings about sources having unknown reputations may have a small backfire effect on trust in accurate sources~\citep{williams-ceci2024misinformation}.
However, aside from brief blog posts announcing the rollout of Google's warning banners, little is known about when or why they're deployed in practice. 

In this study, we develop methods for discovering and evaluating Google's warning banners, build deep learning models to predict the presence of those banners, and use those models to identify unlabeled data voids in web search.
To do so, we used a diverse set of 1.4M unique search queries that were shared on social media.
We obtained these search queries by collecting a larger dataset of \textit{search directives}, which are defined as attempts to prompt people into conducting an online search---e.g., a social media post that tells viewers to search for ``vaccines cause autism''---and have been shown to lead to data voids of low-quality results~\citep{robertson2023identifying}.
Using best practices developed in algorithm auditing studies~\citep{metaxa2021auditing}, we collected the first Search Engine Results Page (SERP) on Google for each of our queries using automated requests from a fixed location, extracted features from those SERPs (e.g., web domains), and merged those features with metrics validated in past work (e.g., the domain quality scores from \cite{lin2023high}).
To account for changes in Google's search results and warning banner systems over time~\citep{munger2019limited}, we collected the search results for all 1.4M queries in three waves that were conducted about five months apart, in October 2023, March 2024, and September 2024.

Across all three data collection waves, about 1\% of the 1.4M search queries produced a warning banner of any type when searched on Google.
Among Google's three distinct warning banners (Figure~\ref{fig:banner_ex}), low-relevance banners were the most common, accounting for 98--99\% of the banners we observed in any wave.
In contrast, and in line with their time-sensitive nature, rapidly-changing banners were the least common, and only 2-10 queries produced one in any wave.
Low-quality banners accounted for 2.1\% of all banners (0.021\% of all queries) seen in October 2023, 1.5\% of all banners seen in March 2024 (0.015\% of all queries), and never appeared in September 2024. 
Though there was substantial churn in the search results between each crawl, with only 37--38\% of the URLs returned at one time step still appearing for the same query about five months later, we found that average domain quality only improved about 0.3\% between crawls.
These results suggest that Google may have discontinued their use of low-quality warning banners around August 2024, despite the quality of their search results remaining largely unchanged.

When low-quality banners did appear, they were consistently more likely to appear on SERPs with lower domain quality scores and for queries containing conspiracy-related keywords.
Low-quality banners also never appeared for search queries containing an advanced query operator, such as ```deep state' site:infowars.com,'' where the site operator (``site:infowars.com'') limits the results for that search to webpages only from that domain.
This absence may be somewhat surprising, especially given the presence of queries in our dataset that contained an advanced query operator to effectively guarantee a data void by filtering the search results to come from a domain with an extremely low quality score (e.g., ``covid vaccine detox site:naturalnews.com'', which has a quality score of 0).
This apparent loophole may follow from the non-committal language used in the banner itself (``some of [these results] may not have reliable information on this topic''); if a low-quality banner appeared on a SERP that was produced by a query with a single ``site:'' operator, then there would be no ambiguity as to which website triggered it.

Before the low-quality banners stopped appearing, we gathered a complementary dataset of SERPs using the subset of queries that produced such a banner in our October 2023 wave.
We collected this dataset in June 2024 using a more rapid collection schedule, with searches conducted roughly four hours apart, for 73 time steps.
To assess the stability of the low-quality banners, we checked for differences in the set of queries that produced a low-quality banner at each time step, and whether differences in the search results returned at each time step might explain the presence or absence of that banner.
On average, 3.2\% of queries that had a low-quality banner in one time step no longer had one when conducted again four hours later, and we found no successive time steps in which the exact same subset of queries received a low-quality banner.
We also could not find a determinative rule for the presence of low-quality banners using the presence and ranking of certain URLs on a SERP, suggesting substantial inconsistencies or randomness in the system that placed the low-quality warning banners.

To identify data voids beyond Google's classifications, we fine-tuned three deep learning models---DistilBERT, a homogeneous Graph Neural Network (GNN), and a heterogeneous GNN---to predict the presence of a low-quality banner given a search query and details about its corresponding SERP. 
We found that the heterogeneous GNN outperformed the other models on a comprehensive set of evaluations, including standard metrics, annotated precision@K, out-of-sample testing, and correlations between the models' predictions and the corresponding SERP domain quality scores.
Using this model, we classified between 0.44\% and 1.16\% of all SERPs as low-quality data voids, which is 29 to 58 times more SERPs than Google placed a low-quality banner on.
These results are consistent with a simpler data voids definition, using an average domain quality cutoff, which identified between 0.63\% and 0.83\% SERPs in our data as low-quality data voids.

Our findings shed light on an important yet understudied content moderation system---Google's warning banners---and highlight the challenges of conducting such studies when the system of interest is a moving target that can frequently undergo rapid, unannounced, and substantive changes~\citep{bagchi2024social}.
One potential explanation for the sudden absence of the low-quality banners is that Google's August 2024 ``core updates''~\citep{aug2024update} improved the quality of its search results such that the low-quality banners were no longer needed.
However, comparing across our main data collection waves and finer-grain datasets, we found little evidence of substantive changes to the domain quality of the SERPs produced for queries that had previously received a low-quality banner.
Instead, we found that many of those queries continued to return low-quality domains in September 2024, they just no longer received a low-quality warning banner.

Discontinuing these warning banners, without replacement or substantial improvements in domain quality, would be a surprising change to make in the months preceding the 2024 US elections.
While we do not measure the impact of that change in this study, events with breaking news updates provide fertile grounds for data voids~\citep{golebiewski2019data}, the use of search directives is not new to political campaigns (e.g., ``Google Ron Paul'';~\cite{baker2008google}), and emerging technologies have made it easier than ever to create misleading content~\citep{feuerriegel2023research}, all of which suggest that keeping the low-quality banners would have been helpful to users.
As such, our findings highlight the need for greater transparency around Google's warning banners, their prevalence, and the effects that their presence or absence has on real users.
Given that we focus on Google Search and use a predominantly English sample of search queries, our results may underestimate the prevalence of data voids, and future work is needed to evaluate them using a broader set of languages, and on a broader set of search engines, including newer LLM-powered ones.
Last, our study also highlights the need for long-term longitudinal studies of online platforms more widely, as changes like the one we observed may otherwise go unnoticed.

\newpage

\section{Results}

\subsection{Evaluating Warning Banner Prevalence and Characteristics}
\label{sec:evaluating}

\para{Warning banners are rare} 
Using our dataset of 1.4M unique search queries that were shared on social media (Methods~\ref{sec:methods-directives}), we collected the corresponding SERP for each query in three primary waves (October 2023, March 2024, and September 2024; Methods~\ref{sec:methods-serps}).
Across all waves, we found that Google displayed one of its three distinct banner types (Figure~\ref{fig:banner_ex}) in the search results for about 1\% of our queries.
Low-relevance banners were the most common banner type, accounting for 97.9\% or more of the banners seen in each crawl.
Low-quality banners were the next most prevalent, accounting for 2.1\% of all banners in crawl-1 and 1.5\% in crawl-2, but never appeared in crawl-3.
In contrast, rapidly-changing banners consistently appeared across crawls but were exceedingly rare, appearing only twice in crawl-1 and crawl-2 (for different queries), and 10 times in crawl-3.
Given how rare rapidly-changing banners were, which was expected due to their time-sensitive nature, our analysis focuses on the low-relevance and low-quality banners.
Additional details on the warning banner types we observed are available in Appendix~\ref{sec:appendix-banners}.

\begin{figure}[t!]
    \centering
    \begin{subfigure}[t]{0.49\textwidth}
        \centering
        \includegraphics[width=\textwidth,trim={0.6cm 0 0.2cm 0.1cm},clip]{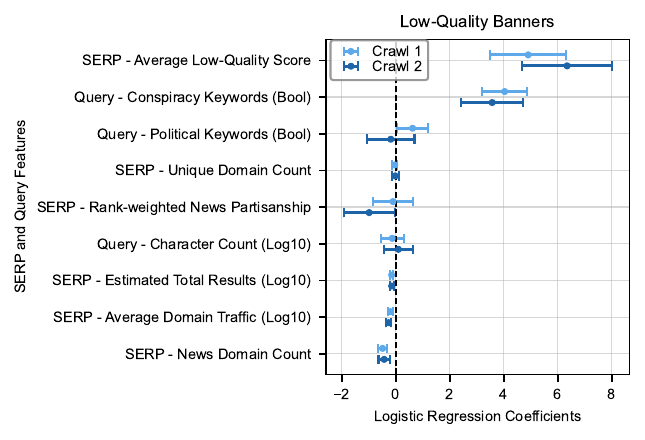}
        \label{fig:logit-quality}
    \end{subfigure}
    \hfill
    \begin{subfigure}[t]{0.49\textwidth}
        \centering
        \includegraphics[width=\textwidth,trim={0.6cm 0 0.2cm 0.1cm},clip]{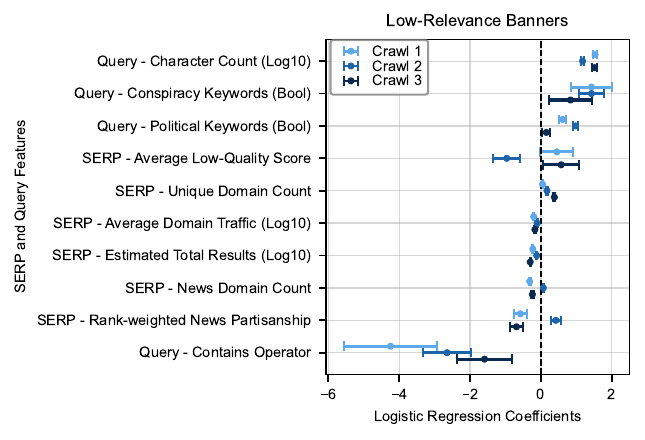}
        \label{fig:logit-relevance}
    \end{subfigure}
    \caption{Across crawls, having low-quality domains in the SERP was the feature most associated with the presence of low-quality banners (left), while the length of a search query was the feature most associated with low-relevance banners (right).
    For both banner types, the use of conspiracy-related keywords in the search query was consistently the second largest positive association.
    In contrast, Google's estimated total number of results, average domain traffic, and the number of news domains on a SERP all had a consistent negative association with the presence of either banner type.
    The use of an advanced operator in a search query had a consistent strong negative association with low-relevance banners, but queries containing such operators never produced a low-quality banner.
    Detailed regression tables are available in Appendix~\ref{sec:appendix-logit}.}
    \label{fig:logit}
\end{figure}

\para{Warning banners are associated with low-quality domains, longer queries, and conspiracy-related queries} 
To evaluate the factors associated with warning banner presence, we used a set of features related to each search query (Methods~\ref{sec:methods-queries}), its corresponding search results, and several domain-level metrics (Methods~\ref{sec:methods-domains}), as dependent variables in logistic regression models (Methods~\ref{sec:methods-logit}).
This includes the presence of partisan or conspiracy-related terms in the search query, and quality scores for the domains appearing in the search results.
In line with its stated purpose, having a low average domain quality score (at the SERP level) was the feature most associated with the presence of low-quality banners.
For low-relevance banners, having longer search queries (word count) was the feature most associated with their presence (Figure~\ref{fig:logit}), which also aligns with their stated purpose, as finding relevant results for long queries may be a challenging task and Google has a 32-word limit on query length (Methods~\ref{sec:methods-directives-queries}).

For both banner types, the use of conspiracy-related keywords in the search query was consistently the second largest positive association with banner presence, and political keywords had a consistent positive association for low-relevance but not low-quality banners.
With respect to negative associations, the estimated total number of results---Google's self-reported estimate for the number of webpage matches found for the query across their full index---and average domain traffic both consistently had a negative association with the presence of either banner type across all crawls.
In contrast to its relationship with low-quality banners, low domain quality had a directionally inconsistent association with the presence of low-relevance banners.

\para{Low-quality banners do not appear when queries contain advanced operators} 
In total, 1.5\% of our search queries contained an advanced operator (Methods~\ref{sec:methods-queries-operators}).
These can be used, for example, to restrict one's search results to a specific website (e.g., ``site:cnn.com''), such that the results only consist of matches for that query within that website.
While our logistic regression models suggest that the use of an advanced operator consistently had a strong negative association with the presence of low-relevance banners across crawls, none of our queries containing such operators ever returned a low-quality banner.
Examining the queries with advanced operators that returned low-quality SERPs (average domain quality < 0.5), we found a number of cases in which the advanced operator was designed to limit the search results to one or more domains with low domain quality scores.
For example, the queries ``ginko site:naturalnews.com'' (which has a quality score of 0), ``site:stormfront.org "sleepy eyes"'' (quality score 0.05), and ``"deep state" site:infowars.com'' (quality score 0.05), all returned search results exclusively from those domains, but never produced a low-quality banner.
We also found more complex queries that combine advanced operators to exert greater control over the results returned (see Methods~\ref{sec:methods-queries-operators}).
In the context of people being intentionally guided into data voids~(\cite{tripodi2023your}; \cite{robertson2023identifying}), these results suggest that the use of advanced operators may provide a loophole for evading Google's warning banners. 

\subsection{Measuring Warning Banner Consistency and Stability}
\label{sec:measuring}

\para{There is substantial churn in low-quality banner presence over time, but small improvements to average domain quality}
Between our first two crawls, we found a decrease in the number of queries that received a low-quality banner (from 0.021\% of all queries in crawl-1 to 0.015\% in crawl-2), and substantial churn in which queries received a low-quality banner.
Among the 301 queries that received a low-quality banner in the first crawl, 154 no longer received one in crawl-2, and 74 queries that had not previously received a banner then did.
These changes were accompanied by substantial churn in the search results returned for our queries, and only 35.7\% (SD=18.0\%) of the URLs returned for a given query in October 2023 were still returned when we searched the same query in March 2024.
We found a similar rate of turnover the gap between the crawl-2 and crawl-3, with only 33.7\% (SD=17.9\%) of the URLs returned in March 2024 still present in September 2024.
Despite this high rate of change in URLs, the average domain quality of the SERPs we collected only improved by 1.0\% overall, from crawl-1 (0.782) to crawl-3 (0.792), and by 0.5\% (SD=6.4\%) when using query-paired differences.
These results suggest that the three core updates and two spam updates that Google completed during our study~\citep{google2025google} did not have a large impact on average domain quality, but these overall averages may hide differences among specific subsets of queries, such as those that received a low-quality warning banner in one of the first two crawls.

\begin{figure}[t!]
    \centering
    \includegraphics[width=\textwidth]{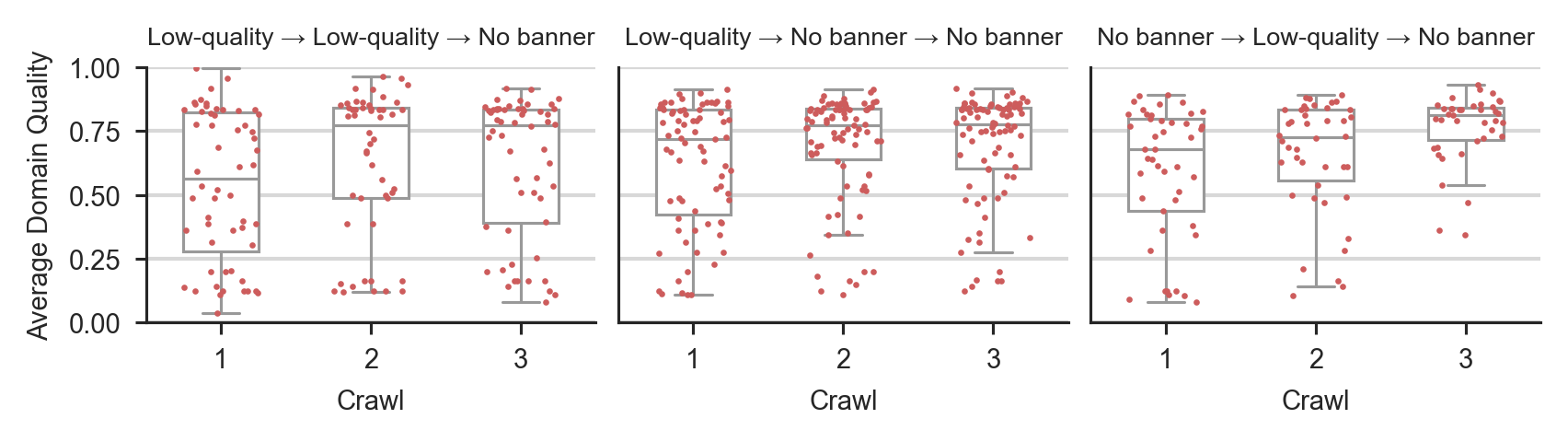}
    \caption{Many of the queries that received a low-quality banner in either (or both) of the first two crawls, continued to produce low-quality SERPs in the third crawl without a banner. Here we present the domain quality score distributions for the top three banner histories, which account for 85.9\% of all banner histories that include a low-quality banner.}
    \label{fig:banner_change_quality}
\end{figure}

\para{Queries that used to receive low-quality banners still return low quality SERPs}
To consider the changes that might have led to the absence of the low-quality banners in crawl-3, we examined the subset of queries that received such a banner in either of the first two crawls (N=375, 0.03\% of all queries).
Although they no longer received a low-quality banner in crawl-3, many of these queries continued to produce low-quality SERPs (Figure~\ref{fig:banner_change_quality}).
For example, among these 375 queries, 34.7\% received a low-quality banner in both of the first two crawls but no banner in the third crawl.
Examining this subset, we found an increase in average domain quality between the first two crawls (0.035 paired, from 0.538 to 0.627) when the banner was present, and a smaller decrease between the second and third crawls when the banner no longer appeared (0.022 paired; from  0.627 to 0.619).
A similar pattern held for the 33.3\% of these queries that received a low-quality banner in the first crawl but no banner in either of the subsequent crawls.
Last, for the 17.9\% of these queries that received a low-quality banner in the second crawl, but no banner in the first or third crawls, we found that their average domain quality scores consistently increased across crawls (0.077 paired crawl-1 to crawl-2, 0.072 paired crawl-2 to crawl-3).
These results suggest that, rather than no longer being needed due to a shift in quality, the low-quality banners may have been discontinued.

One of the queries that received a low-quality banner in both crawls 1 and 2, but not 3, is ``naturalnews "the coming delta lockdown is designed to invoke nationwide protests".''
This query consistently returned the same exact webpage in top ranked result---a naturalnews.com article that claims the COVID lockdowns for the delta strain were ``designed to invoke nationwide protests ... to blame "anti-vaxxers"''---and other results that generally support a similar narrative.
Another example is ``vril lizards droning process,'' which had more turnover in the results, but ultimately returned a similar set of websites supporting a conspiracy theory about mythical parasite lizards (see Appendix~\ref{fig:appendix-vril}). 
Last, the query ``"underground war" qanon blessed2teach'' returned results leading to QAnon-related content in every crawl, and searching this query at the time of this writing (February 2025), the top result leads to a webpage titled ``Is Trump Safe? CIA, MI6, Mossad Plot ...''

\begin{figure}[tb!]
    \centering
    \begin{subfigure}[t]{0.49\textwidth}
        \centering
        \includegraphics[width=\textwidth]{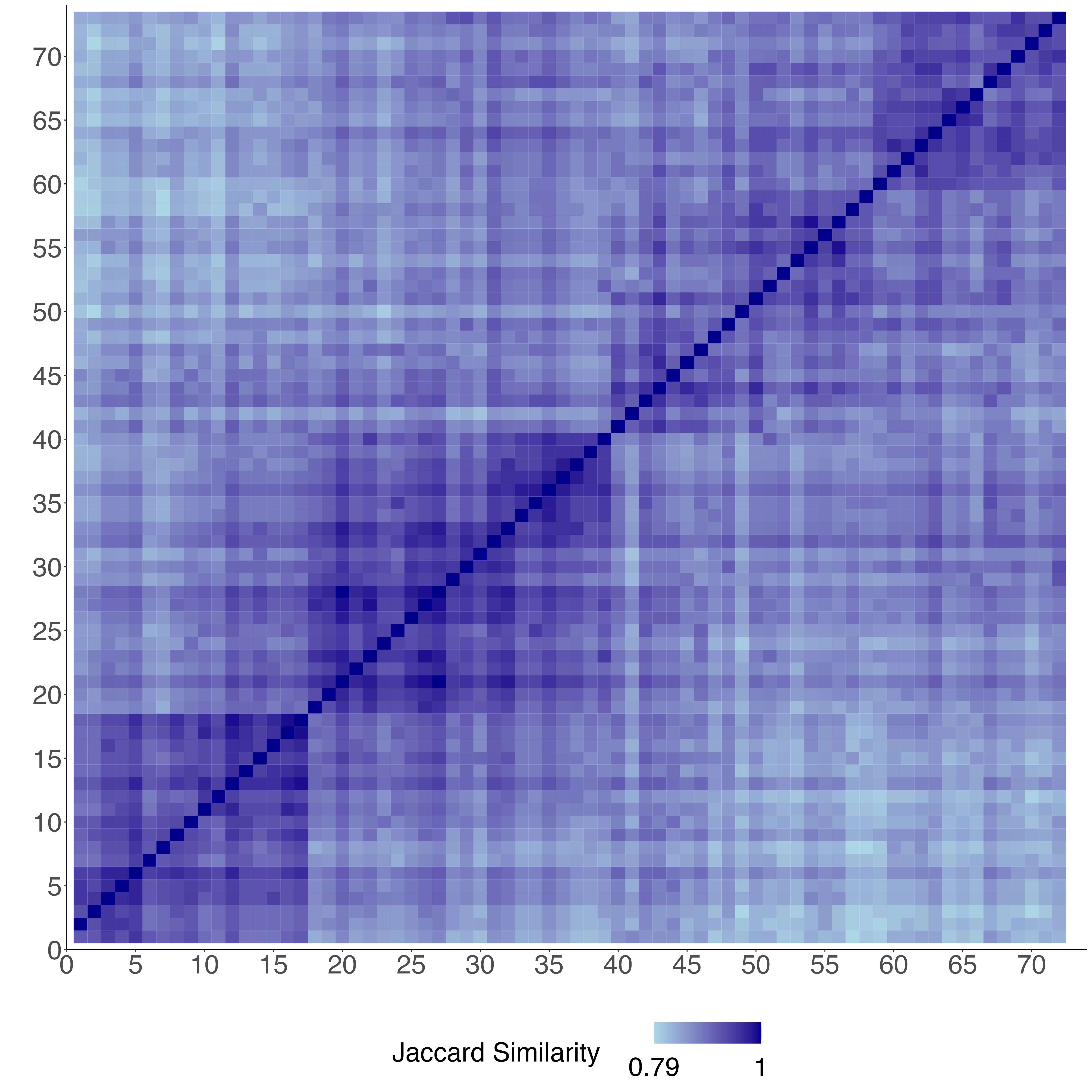}
        \caption{Heatmap cells show the pairwise Jaccard similarity of the set of queries that returned a quality banner over all 73 time-steps. Data collection issues resulted in gaps at steps 18 and 41, which account for the steepest plot gradient changes (see Appendix~\ref{sec:appendix-consistency}).}
        \label{fig:jaccard_qry}
    \end{subfigure}
    \hfill %
    \begin{subfigure}[t]{0.49\textwidth}
        \centering
        \includegraphics[width=\textwidth]{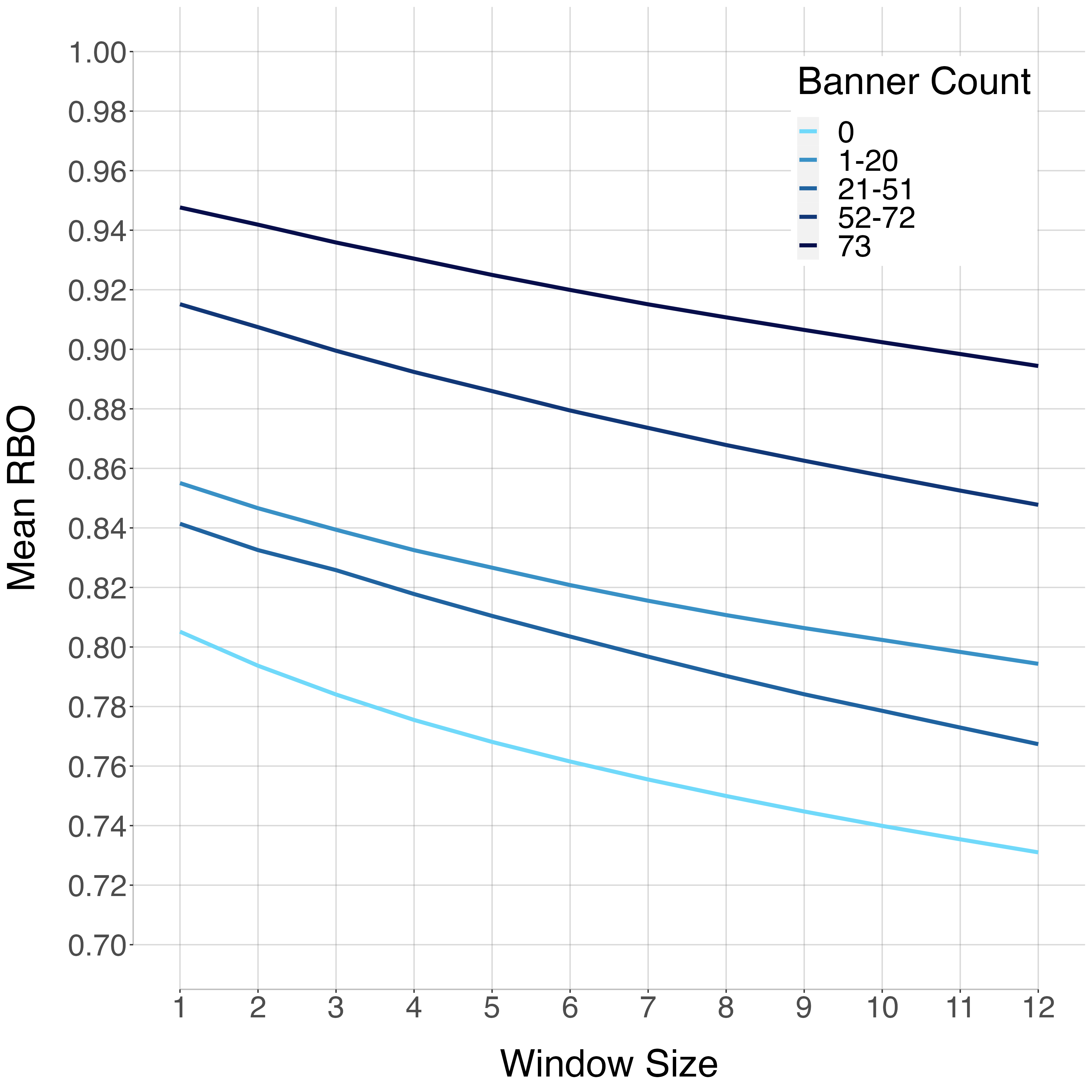}
        \caption{Similarity of SERPs measured by using mean \(RBO_k\) (y-axis) over 12 window sizes (x-axis). Queries were divided into groups based on whether the SERPs they produced received a banner in 0, 1-20, 21-51, 52-72, or all 73 time steps (legend).
        }
        \label{fig:rbogroups}
    \end{subfigure}
    \caption{The set of queries that receives a low-quality banner frequently change over short time spans, and queries that consistently receive banners also have relatively consistent search results.}
\end{figure}

\para{Low-quality banner presence is inconsistent over short time spans} 
Prior to the discontinuation of the low-quality banners, we collected a temporally dense set of search results using the 301 queries that received a low-quality banner in the first crawl. 
We conducted these queries on a more rapid collection schedule for a period of 73 time steps spaced about 4.5 hours apart (Methods~\ref{sec:methods-stability}), resulting in 1.05M search results over 22K SERPs. We collected the first 100 results for each query, but many queries consistently had fewer than 100 results. Using the Jaccard index as a pairwise measure of similarity between each timestep, we found that the minimum similarity between any two time periods was 0.79, the average similarity (excluding the diagonal) was 0.88 (Figure~\ref{fig:jaccard_qry}), and there was only one instance in which two time steps shared the exact same set of queries.
These results are consistent with a similar dataset that we collected in March 2024, where we used the same set of 301 queries, but collected data every 1.5 hours for 34 time steps (see Appendix~\ref{sec:appendix-consistency-pilot}).
These results suggest that Google's policy for placing low-quality banners is not based solely on text of the search query, turning our attention to the distribution of results returned for a query.

\para{Low-quality banner presence is more consistent for stable results} 
To examine differences over time, we used Rank-Biased Overlap (RBO), a similarity metric well suited for comparing ranked lists~\citep{webber2010similarity}, with a sliding window ($RBO_k$; see Methods~\ref{sec:methods-stability}).
We found that the queries that consistently received a low-quality banner in the temporally dense dataset also had the most consistent search results over time (Figure~\ref{fig:rbogroups}). 
Among the 301 queries used to construct this dataset, five did not return search results at any time step, 166 did not produce a low-quality banner at any timestep, and 40 always produced one, leaving 90 with at least some variation in their banner production.
To help account for these cases, we calculated $RBO_k$ for several different banner-count groups, but found that RBO monotonically decreases as window size $k$ increases across all groups, highlighting a consistent churn in search results over time.
However, despite sharing this general pattern, the queries that always returned low-quality banners had the most stable search results across all window sizes, and the queries that did not produce any low-quality banners had the least stable search results at every $k$. 
Consistent with our logistic regressions on the main crawl data, these results suggest that the search results, rather than the search query, may be the more important factor for determining why a low-quality banner was shown.

\begin{figure}[tb!]
    \centering
    \begin{subfigure}[t]{0.49\textwidth}
    \centering
    \includegraphics[width=\textwidth]{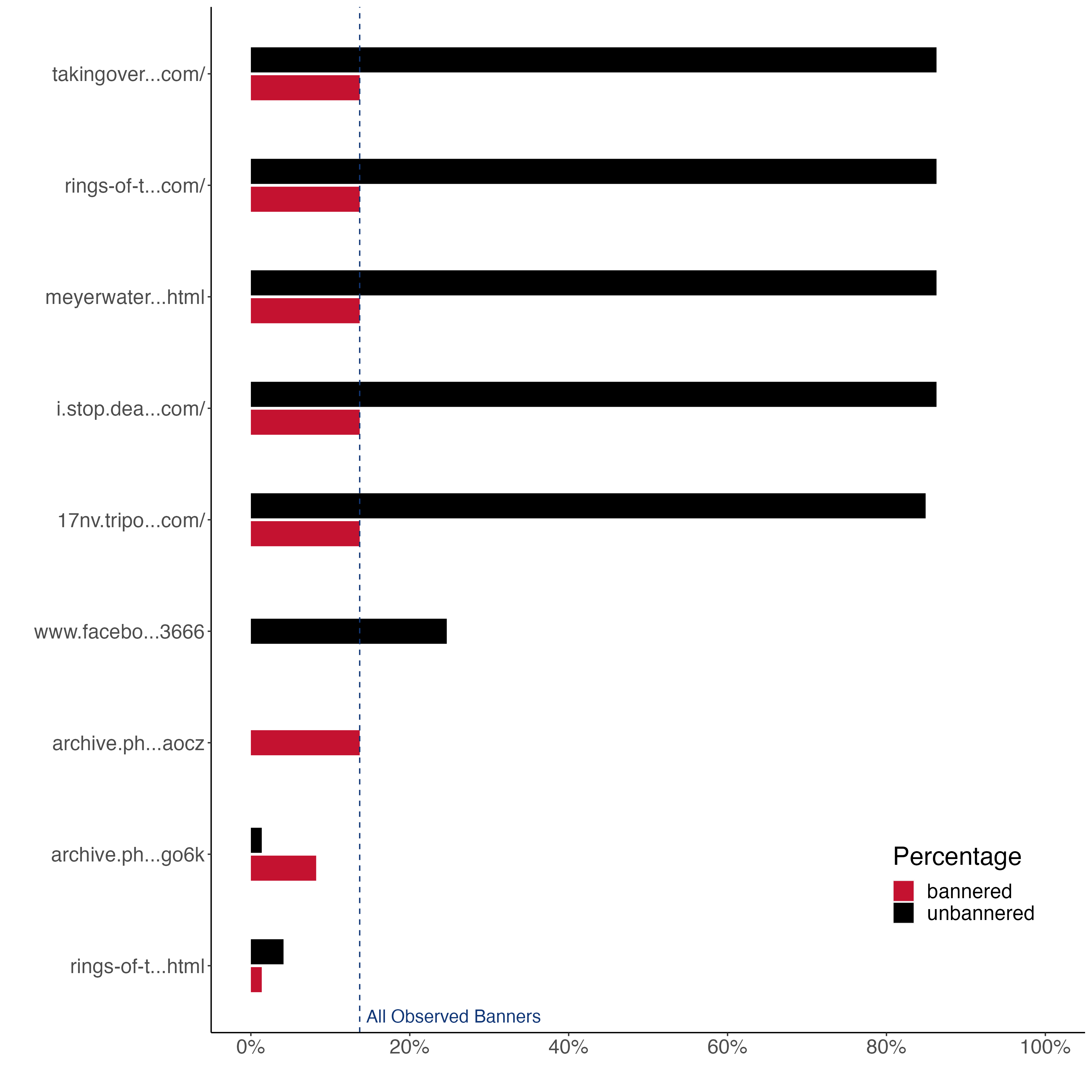}
    \caption{The presence of the first archive.ph URL (``archive.ph...aocz'') can explain all banners displayed for ``advanced search result manipulation technology'' observed over 73 time steps.}
    \label{fig:cip_bar}
    \end{subfigure}
    \hfill
    \begin{subfigure}[t]{0.49\textwidth}
    \centering
    \includegraphics[width=\textwidth]{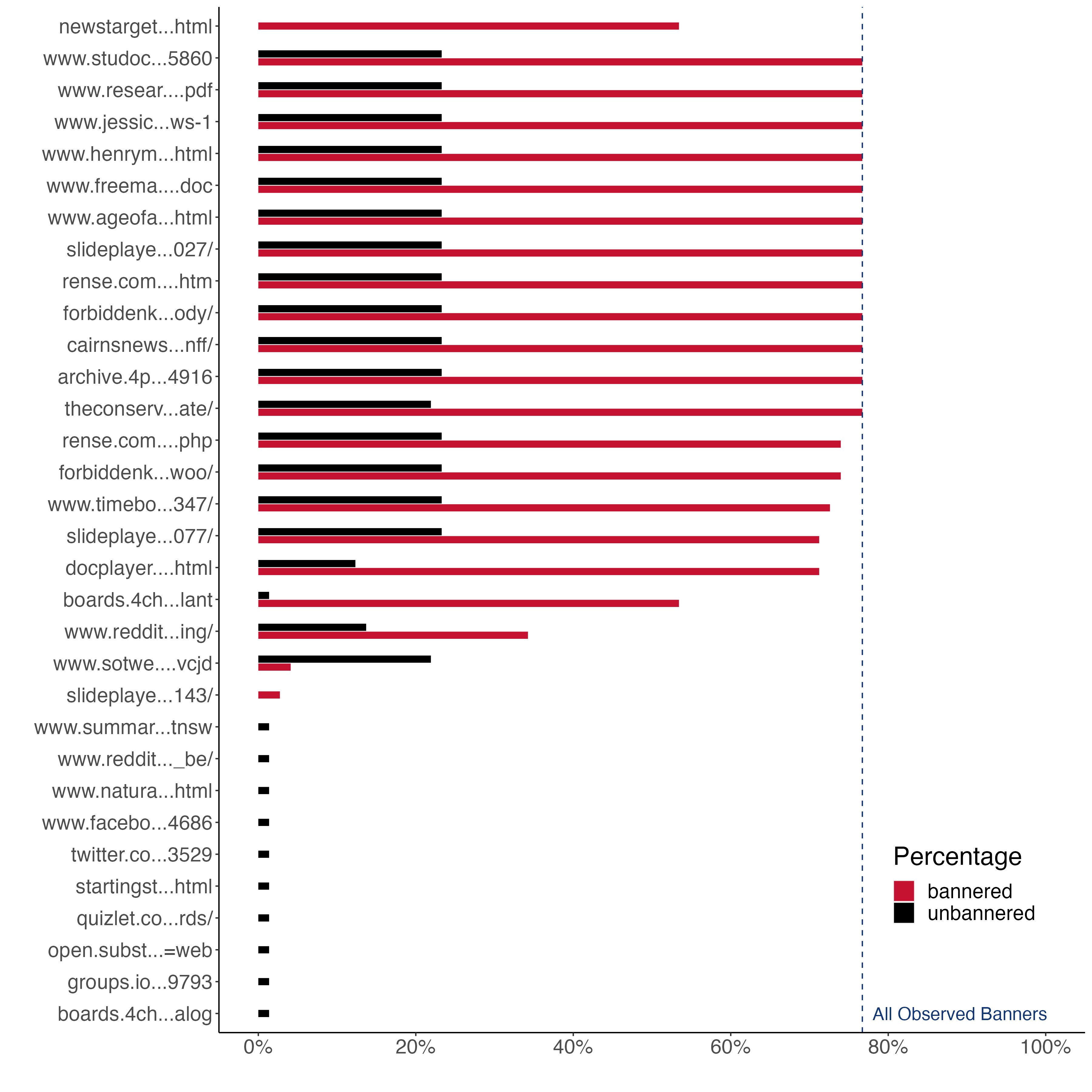}
    \caption{While some URLs are more associated with banners than others, no single URL can explain every banner for ``mrna prions''.}
    \label{fig:mrnaq1}
    \end{subfigure}
    \caption{The presence of specific URLs can sometimes fully explain the presence of low-quality banners for certain queries (left), but not for others (right). The vertical line indicates the total number of banners observed for the query. If every time a URL appears there is a banner (red bar) and never appears with no banner (black bar), the URL can fully explain observed banners for the query.}
\end{figure}

\para{The presence of specific URLs does not fully explain low-quality banner presence} 
To examine whether the presence of certain URLs might trigger low-quality banners, we used the subset of data for the 90 queries in our temporally dense dataset that produced SERPs both with and without low-quality banners.
We use this subset of queries because the variance in their banner presence can be leveraged to compare the conditions under which a banner does or does not appear (Appendix~\ref{sec:methods-stability-dependency}). 
For 25 of these 90 queries, we were able to pinpoint a single URL that was always present in its search results when it received a low-quality banner, but never present when it did not (Figure~\ref{fig:cip_bar}).
However, the remaining majority of queries did not have a single URL that met the same conditions.
For example, the query \textit{``mrna prions''}---which consistently surfaced anti-vaccine articles and social media posts in its top 10 results (Appendix \ref{sec:appendix-examples-prions})---returned a low-quality banner in 56 of the 73 timesteps and had a URL that was highly associated with banner presence, but its presence in the SERP was not required for the query to receive a banner (Figure \ref{fig:mrnaq1}).
Extending this measure to account for the presence of two URLs only helped to explain low-quality banner presence for four additional queries (29 of the 90 queries). 
Further extending this measure to three URLs allowed us to fully explain 67 of the 90 queries.

\para{URL rankings do not fully explain low-quality banner presence}
We further examined the role of URL presence by conditioning URL-pairs on rank cutoffs.
This approach allowed us to check for determinative cases where a pair of URLs appeared in the search results, within a specific range of rankings, and always received a banner (see Appendix~\ref{sec:appendix-consistency} for a formal definition).
We found that at least one rank cutoff could explain the presence of low-quality banners for 65 queries of 90 across all timesteps, and a cutoff of $c=1$ (i.e., the top-ranked URL) was able to explain the largest proportion of those queries (48 of 65). 
Although some queries were fully explainable with larger cutoff values (i.e. $c>1$), the proportion of banners explainable by subsequent cutoffs ($c=\{2, \dots, 50\}$) decreases monotonically.
These results suggest that the first results on a SERP are an important factor in determining banner placement, but still do not fully explain the presence of low-quality banners.

\subsection{Predicting Warning Banner Presence}
\label{sec:predicting}

Advancing on our evaluation of the policies that govern Google's warning banners, we also used our dataset to develop deep learning models that can incorporate additional context and potentially aid in proactively identifying data voids. 
To that end, we built and tested three models: a homogeneous Graph Neural Network (GNN) and a heterogeneous GNN trained on a bipartite query to domain graph, and a fine-tuned DistilBERT model trained on query text (see Methods \ref{sec:methods-models}). Domain features are constructed only from content on SERP pages, and contain no domain-level features or domain reliability annotations. We demonstrate that even a relatively small GNNs with 1.05M parameters can effectively identify data voids and surface SERPs associated with unreliable domains. While we already demonstrated that query text alone cannot explain all banners, the association of quality banners with conspiratorial terms suggests some linguistic patterns may be present. We therefore elect to include DistilBERT as a text-only baseline. Given the small set of low-quality banners we observed, and the propensity of deep learning models to overfit, we include multiple evaluations that demonstrate the effectiveness of our proposed models.

\para{The GNN models outperform DistilBERT}
In line with our finding that low-quality banners do not depend on the text of the search query text alone, the DistilBERT (text-only) baseline was outperformed by the GNN approaches on all metrics except precision (Table~\ref{tbl:GNNResults}).
Among the GNN models, the heterogeneous GNN performed the best, outperforming both DistilBERT and the homogeneous GNN on accuracy, F1, precision, and recall. 
To calculate these statistics, we trained and evaluated each model ten times, each time drawing a different negative sample for the non-bannered class to mitigate the possibility of overfitting to a single negative sample. 
Additional details on these models, including preprocessing and model design, are available in Methods~\ref{sec:methods-models}.

\begin{table}[t!]
    \small
    \centering
    \caption{GNNs perform best on hold-out data and best identify new out-of-sample queries with low-quality banners. This table provides the mean and standard deviation for evaluation metrics over 10 runs on the hold-out test-set (left), and the out-of-sample evaluation (right), which includes the number of queries (out of the 74 queries that received a low-quality banner in crawl-2 but not crawl-1) in each model's $K$ most confident predictions.}
    \label{tbl:GNNResults}
    \begin{tabular}{rllllccc}
\toprule
& \multicolumn{4}{c}{Test Set Evaluation} & \multicolumn{3}{c}{Out-of-Sample Evaluation} \\
\cmidrule(lr){2-5} \cmidrule(lr){6-8}
Model & Accuracy & F1 & Recall & Precision & Top 100 & Top 500 & Top 1k \\
\midrule
DistilBERT  & 0.87 $\pm$ .01          & 0.68 $\pm$ .03          & 0.57 $\pm$ .04 & 0.84 $\pm$ .07 & 2 & 9 & 12 \\
GNN\textsubscript{Hom}   & 0.90 $\pm$ .00          & 0.87 $\pm$ .01          & 0.72 $\pm$ .02 & 0.88 $\pm$ .00 & 5 & \textbf{13} & \textbf{16} \\
GNN\textsubscript{Het}   & \textbf{0.93 $\pm$ .01} & \textbf{0.90 $\pm$ .01} & \textbf{0.78 $\pm$ .04} & \textbf{0.92} $\pm$ .00 & \textbf{6} & 11 & 15 \\
\bottomrule
\end{tabular}

\end{table}

\para{GNNs best identify out-of-sample low-quality banners} 
We assessed our models' out-of-sample performance by using the differences in low-quality banner presence between crawl-1 and crawl-2.
Specifically, we examined how well each model confidently identified the 74 queries that did not have a low-quality banner in crawl-1 (October 2024), but did have one in crawl-2 (March 2024). 
To do so, we assessed how many of these 74 queries our models detected within their $K$ most confident quality banner predictions (with labeled data excluded).
The GNN models again outperformed the DistilBERT model, with the homogeneous model correctly including more of the newly-bannered 74 queries in its most confident 50, 500, and 1K predictions, and the heterogeneous model correctly including more of these queries in its most confident 100, 5K, and 10K predictions (Table~\ref{tbl:GNNResults}; see Appendix~\ref{sec:appendix-gnn} for additional details).
While these tests help establish the utility and validity of our GNN models~\citep{inoue2005sample}, we note that the inconsistencies we observed in low-quality banner placement over short time periods (Section~\ref{sec:measuring}) means the set of 74 queries and their corresponding SERPs that we used are not necessarily reflective of those that should receive a low-quality banner.

\begin{figure}[tb!]
\centering
\includegraphics[width=0.9\textwidth]{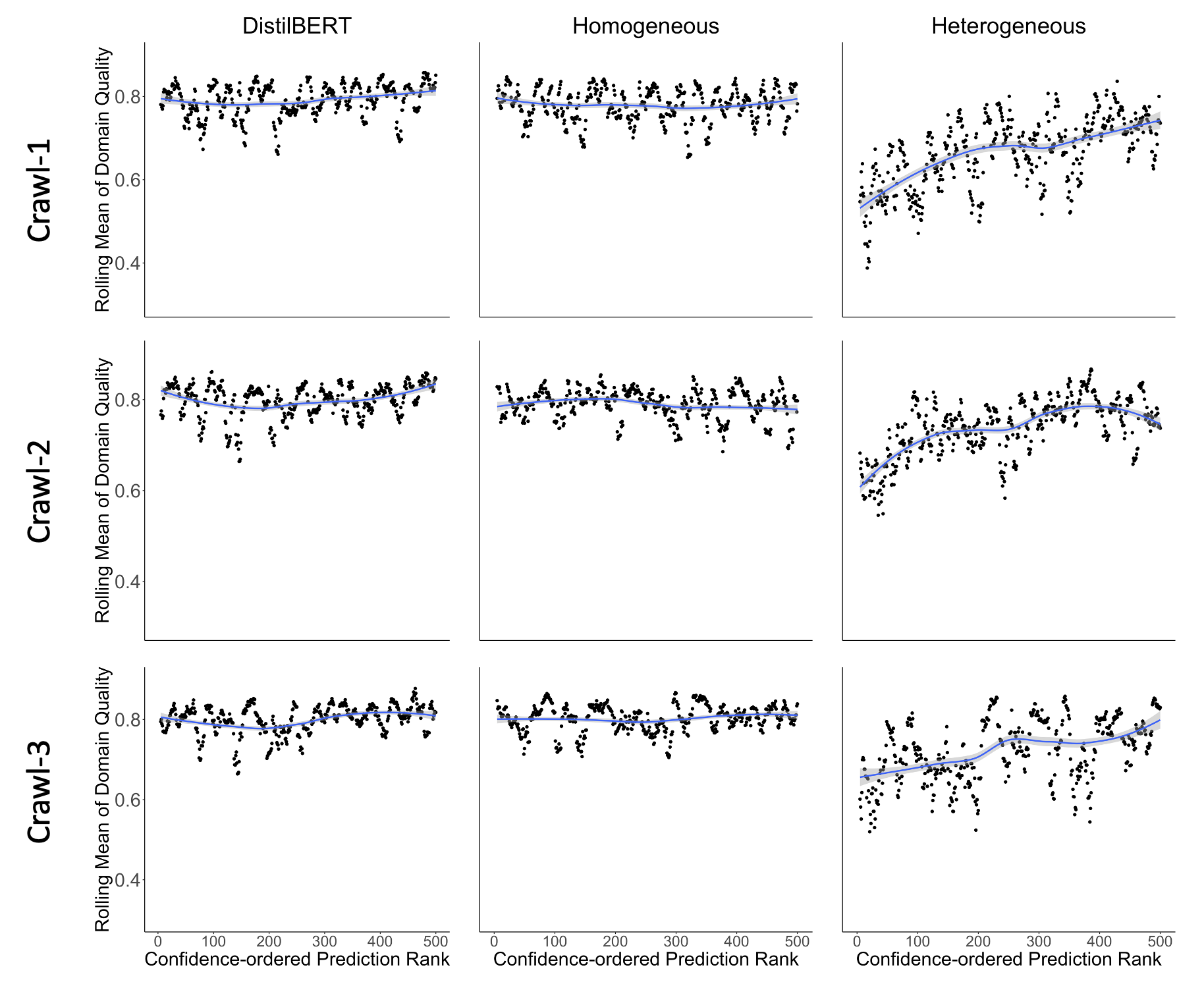}
\caption{Heterogeneous GNNs best identify SERPs associated with unreliable domains. Each graph displays the rolling mean (window size of 10) of Domain Quality Scores over the 500 highest-confidence model predictions (lower scores indicate less reliable websites). With models trained on Crawl-1, we evaluated quality banner predictions for Crawl-1 (top), Crawl-2 (middle), and Crawl-3 (bottom).}
\label{fig:qualwindow}
\end{figure}

\para{GNNs identify SERPs with low average domain quality} 
Given the focus on reliability in the stated purpose of the low-quality banners (Figure~\ref{fig:banner_ex}A), the prevalence of domains with low-quality scores in a query's SERP provides a useful proxy for model evaluation.
A model's most confident ``low-quality banner'' predictions should be associated, on average, with the presence of less-reliable domains.
To investigate this relationship, we used the rolling mean of average domain quality over the 500 most confident banner predictions obtained from each model, and found the heterogeneous GNN model's predictions had the strongest relationship with domain quality (Figure~\ref{fig:qualwindow}). 
Average domain quality was calculated for each query as the simple average over all of its returned search results in each crawl, and we used a rolling mean with a window size of 10 to account for inherent noise in this metric.
Across crawls 1, 2, and 3, the heterogeneous model displayed the most robustness, and continued to identify queries associated with low-reliability domains on data extracted 5 months after its training data. As only 890k SERPs in crawl 1 had at least one non-social media website with a domain quality label,
many predictions across models are excluded. We note that the heterogeneous model is able to confidently predict SERPs associated with lower-quality domains despite our models not including any explicit domain-level reliability features.
These results suggest that the heterogeneous GNN model best learned to identify queries associated with unreliable domains, which helps validate our findings, and again point to the heterogeneous model as the best performer.

\para{GNNs identify queries that annotators agreed should have low-quality banners}
After manually examining the 301 queries that received a low-quality banner in crawl-1 (October 2023), two annotators labeled the top 20 predictions of each model based on their corresponding SERPs. 
One of these labels indicated whether or not the SERPs produced by each query should receive a low-quality banner (i.e. returned results that appeared to be low-quality), and there was substantial agreement (Cohen's $\kappa = 0.73$) between the two annotators~\citep{landis1977measurement}.
Both annotators agreed that the heterogeneous GNN model yielded the highest precision for the top 5, 10, and 20 predictions (Appendix~\ref{sec:appendix-gnn}, Table~\ref{tbl:precisionCIP}).

\subsection{Proactively Identifying Data Voids}
\label{sec:identifying}

To evaluate the prevalence of data voids on Google Search, we consider defining data voids in three ways.
First, we can consider a data void to be present when a query returns a low-quality banner, as we did in our initial analysis.
Second, we can consider a data void to be present when a query returns a SERP with an average domain quality below 0.5.
Last, we can consider a data void to be present when a query returns a SERP that our $GNN_{het}$ model predicts a low-quality banner for with a high confidence threshold.

\para{There is a disconnect between low-quality SERPs and low-quality banners}
If we define a low-quality SERP as one whose average domain quality score is less than or equal to 0.5 (on a scale of 0--1), then 0.83\% of queries in crawl-1, 0.72\% in crawl-2, and 0.63\% in crawl-3 returned low-quality SERPs.
While this rate is relatively small, it still represents a \(\approx40\times\) increase in SERPs classified as data voids relative to the proportion of SERPs that Google placed a low-quality warning banner on in crawl-1 (0.021\%) and crawl-2 (0.015\%), and continued to identify a similar percentage of SERPs as data voids in crawl-3 when Google displayed zero low-quality banners. 
Using this definition, we found that Google only applied its low-quality banner to 0.45\% of low-quality SERPs in crawl-1 and 0.28\% in crawl-2. 
Put differently, 17.9\% of the SERPs that Google placed a low-quality banner on were low-quality data voids in crawl-1, and 13.1\% in crawl-2.

\para{The number of low-quality banner predictions exceeds observed banner prevalence}
Another approach to defining a data void is to use Google's warning banners as labeled data for building a classifier as we did (Section~\ref{sec:predicting}). 
Using predictions from our $GNN_{het}$ model, which we found to be the most reliable for identifying searches that should have received a banner, we considered its most confident predictions for whether a SERP should receive a low-quality warning banner. 
Here we define a low-quality data void as one which the model predicts a banner for, with a \(\geq 90\%\) confidence score threshold. 
Applied across our dataset, this definition suggests that 1.16\% of SERPs in crawl-1, 0.44\% of SERPs in crawl-2, and 0.72\% of SERPs in crawl-3 should have received a banner.
Similar to our definition of data voids based on an average domain quality threshold, our model's predictions suggest that there were about \(\approx58\times\) more data void SERPs than Google's low-quality banners covered in crawl-1 (about \(\approx29\times\) for crawl-2).
Among those data void SERPs, Google only applied a low-quality banner to 2.0\% in crawl-1 and 2.6\% in crawl-2.
These results suggest that Google's low-quality banners may have been tuned to a very conservative confidence threshold, weighing the consequences of false positives (an unnecessary banner) higher than false negatives (no banner when needed).

\section{Discussion}
\label{sec:discussion}

In this study we developed and deployed methods for surfacing, evaluating, and predicting warning banners and data voids on web search engines.
Using data collected over several waves of varying size and inter-collection latency, we found that Google's low-quality banners were rare, governed by inconsistent and evadable rules, 
and appear to have been discontinued around August 2024.
We do not find that the disappearance of these banners can be explained by improvements in average domain quality, or by replacement with a new banner type, suggesting that the system for placing those banners has either become exceedingly conservative or turned off entirely.
These findings raise questions about the reliability and transparency
of Google's content moderation systems, their capacity and willingness warn users about unreliable information in their search results, and the timing of the changes, which occurred in the months preceding the 2024 US Presidential election.

When Google was using its low-quality warning banners, our results show that it rarely displayed them for the search queries in our sample (0.02\%), and the rules governing their presence on a page of search results were inconsistent and unstable over short time spans.
Our findings also show that these rules can be easily evaded through the use of advanced operators in the search query---as Google never displayed a low-quality banner for such queries---and highlight examples of search queries shared on social media that abuse this loophole to guide people into data voids (e.g., the query ``covid vaccine detox site:naturalnews.com'').
The absence of low-quality banners in such scenarios could be framed as respect for user choice: Google has stated that ``users may decide to seek and select content that our signals determine to be of low-quality ... we believe it is of fundamental importance to respect their choices''~\citep{google2024information}.
While this decision may be reasonable for users formulating their own queries, it ignores the social sharing of search queries with advanced operators, where not all users will understand their impact on the search results.
Alternatively, the non-committal language consistently used to describe the conditions under which low-quality banners are shown may also help explain this loophole.
This includes announcement of the low-quality banner (``This doesn't mean that no helpful information is available, or that a particular result is low-quality''~\cite{nayak2022new}), the text on the banner itself (``some of [the results] may not have reliable information on this topic,'' Figure~\ref{fig:banner_ex}), and the lack of a ``there aren't \textit{any} great results'' banner variant as we found in the low-relevance banners (``there aren't \textit{many/any} great matches,'' see Methods~\ref{sec:methods-queries-operators}).
If a low-quality banner were returned for a search query including a single ``site:'' operator, limiting all results to a single domain, the ambiguity of which website triggered the warning would be removed.

One potential explanation for the absence of low-quality banners since August 2024 is that Google's recent ``core updates''~\citep{aug2024update}, which went into effect around the same time, have improved the quality of its search results such that the low-quality banners are no longer needed.
However, comparing across data collection waves, we find little evidence of substantive increases in the domain quality of the SERPs returned for queries that had previously received a low-quality banner.
Instead, many of those queries continued to return low-quality domains, but no longer receive any type of warning banner.
More importantly, when considering the rate at which Google applied its low-quality banners to different definitions of data voids, including a simple average domain quality threshold and a heterogeneous GNN, we identified up to 50x more SERPs as data voids than Google applied its low-quality banners to. 
If we conservatively extrapolate from Google's self-reported 3.3 billion searches per day in 2012~\citep{google2013google} to 5 billion per day in 2024, and apply the data void rate estimated by our model (0.77\%, averaged across crawls)---which is not representative of the queries real users search but does provide a broad and diverse sample---then the estimated number of data voids encountered per day on Google would be 38.5M. 
Using the percentage of these data voids that received one of Google'’'s low-quality banners in our study (0.3\%, averaged across crawls), we can further estimate that only 115.5K of those 38.5M would have received a low-quality banner.

Our use of search directives to identify a set of search queries is both an advantage and limitation of our study. 
Compared to prior algorithm audits of web search, where researchers often select queries on their own or solicit them from survey participants (\cite{ballatore2015google}; \cite{norocel2023google}; \cite{lurie2021searching}; \cite{vanhoof2022searching}), our use of search directives allowed us to develop one of the largest query sets ever used in an algorithm audit, which was crucial to our goal of surfacing a sufficient sample of data voids and Google's warning banners. 
However, the search directives we used for this purpose come from a long time window (2006 to 2023), so the search results we retrieved for each query do not reflect the results one might have seen had they conducted that search at the time the search directive was posted. 
Similarly, we used queries from search directives that were intended to be used on search engines other than Google, but Google is the largest search engine in the world, and developing tools for collecting and parsing search results from each search engine would have taken a massive infrastructure investment for what is often a moving target.

In addition, we also only examine search directives that were shared as links---which provide a shortcut to search results with pre-specified query---and did not include queries shared in text or images~\citep{robertson2023identifying}. 
Beyond explicit search directives that provide a specific search query to use, users may also be compelled to conduct searches after being exposed to more implicit forms, such as keywords or phrases that are repeatedly promoted by influencers or political elites~\citep{tripodi2019devin}, 
and the recently proposed concept of ``dredge words''~\citep{williams2024dredge}---queries for which unreliable websites rank highly---may offer a promising and scalable avenue for identifying queries likely to produce data voids.
Future work that develops methods for surfacing explicit and implicit search directive queries in real time may help researchers examine the rapidly-changing warning banners that were vanishingly rare in our dataset.

Given Google's public efforts to understand and improve the reliability of their rankings~\citep{google2023search}, our results may be viewed as a lower bound for the prevalence of data voids in web search engines more broadly. 
Other search engines, especially those that are reportedly favored by conspiracy theorists~\citep{thompson2022fed}, may produce a higher rate of data voids,
and should be investigated in future work. 
Similarly, while the queries we searched do not provide a representative sample of real users' queries, only 0.1\% contained a conspiracy-related keyword from a relatively conservative and non-exhaustive list. 
While it is unclear what that proportion would be for real users' queries, their relative scarcity in our dataset also suggests that our results may provide a lower bound estimate for the prevalence of data voids in web search.

The harmful nature of data voids, combined with the potential of search directives to guide people into them, both suggest the prevalence rates we found should provide cause for concern.
Our study sheds light on both the use of warning banners and the prevalence of data voids in web search, but without any official reports on the use or effectiveness of warning banners in search, it remains unclear how harmful the unlabeled data voids that we discovered may be to real users.
Future work should explore independent approaches to collecting exposure and engagement with warning banners among real search engine users~\citep{feal2024introduction}, collaborations with industry that could facilitate that research, and examine a broader set of search engines, languages, and locations~\citep{borge2021how}.

As search evolves and continues to incorporate Large Language Models (LLMs) and agent-based approaches to task completion~\citep{white2024advancing}, studies like ours may become both increasingly important and difficult to conduct. 
The methods we developed here are platform agnostic, but LLMs add greater stochasticity to the outputs of search engines, may change how people write their search queries, and introduce a conversational interface that could change fundamental aspects of how search engines were previously used. 
While recent research suggests that LLMs may be able to reduce conspiracy beliefs when specifically prompted to do so~\citep{costello2024durably}, it's unclear whether popular LLMs will incorporate the prompts needed to elicit that behavior, especially in the context of emerging conspiracies and the incorporation of LLMs into search engines. 
Indeed, with respect to data voids, LLMs may make them easier to create, harder to identify, and more likely to be presented as a valid response to a user's query. 
As such, the intersection of search directives, data voids, and the content moderation practices of both search engines and LLM providers, especially as two become more intertwined, presents an important and pressing area for future research.

\clearpage
\newpage

\section{Methods and Data}
\label{sec:methods}

To conduct this study, we first collected a large dataset of search directives, which are defined as prompts to conduct an online search (Section~\ref{sec:methods-directives}). 
We then used the 1.4M search queries we found in that dataset to surface Google's warning banners (Section~\ref{sec:methods-serps}), evaluate their presence in the context of established query (Section~\ref{sec:methods-queries}) and domain-level metrics (Section~\ref{sec:methods-domains}) with logistic regression (Section~\ref{sec:methods-logit}), test their consistency over time (Section~\ref{sec:methods-stability}), and develop deep learning models to identify unlabeled data voids (Section~\ref{sec:methods-models}).

\subsection{Search Directives}
\label{sec:methods-directives}

We collected search directives from social media posts (Section~\ref{sec:methods-directives-posts}) to collect a diverse set of search queries (Section~\ref{sec:methods-directives-queries}) for our study.
By specifying a flexible linguistic strategy (prompts to conduct an online search) rather than specific content (search queries), search directives provide a useful tool for surfacing unspecified and unknown content.
The search queries used in search directives have been shown to cover a diverse array of topics, ranging from music, sports, and advertising, to medical misinformation about Ivermectin, an emerging conspiracy about the COVID-19 vaccine causing people to ``die suddenly,'' and a cryptocurrency scam~\citep{robertson2023identifying}. 
Although the potential harms of people being led into data voids like these have been well documented~\citep{golebiewski2019data}, few studies have examined how people can be led into data voids (\cite{tripodi2019devin}; \cite{tripodi2023your}), how to computationally identify data voids~\citep{flores-saviaga2022datavoidant}, or how to measure the bridge between social media and search engines more broadly (\cite{bode2018studying};\cite{lukito2020coordinating}; \cite{yarchi2021political}; \cite{zuckerman2021why}).
Rather than on relying on smaller sets of queries generated through surveys or  interviews, or medium-sized sets of queries generated via autocomplete (\cite{robertson2019auditing}; \cite{haak2023qbias}), our use of search directives allowed us to collect 1.4M unique queries without defining the topic space or a starting set of queries to expand upon.

\subsubsection{Social Media Posts}
\label{sec:methods-directives-posts}

We collected a total of 5.25M posts that contained a URL fragment (e.g. \nolinkurl{google.com/search}) leading to one of 25 popular search engines. 
This collection strategy allows for flexibility in subdomains, variability in URL parameters, and enabled us to easily and accurately extract search directive queries.
Following past work, we filtered out URLs that did not lead to a page of search results, including those that did not contain a known query parameter (e.g., ``\&q=\{query\}'' for Google Search) and those that contained a blank query, leaving 4M search directive posts, 4.17M URLs (posts can contain multiple URLs), and 1.44M unique queries that were created by 1.82M unique accounts over a 16.5-year window (2006 to 2023). 
Advancing on prior work that examined the five most popular modern search engines in the US (Google, Bing, DuckDuckGo, Yahoo, and Brave), we used a list of 25 search engines to collect our dataset (Google, Bing, DuckDuckGo, Yahoo, Brave, AOL, Ask, Baidu, Dogpile, Ecosia, Exalead, Excite, Hotbot, Lycos, Metacrawler, Mojeek, Petalsearch, Qwant, Sogou, Startpage, Swisscows, Webcrawler, Yandex, You, and Youdao), including search engines that are prominent outside of the US (e.g. Yandex), were prominent in the past (e.g. AOL and Ask), or newer search engines that feature large language models (e.g. You.com).

\subsubsection{Search Directive Queries}
\label{sec:methods-directives-queries}

Of the 4.17M posts that contained a URL fragment and a search query---which excludes links to search engine homepages that don't qualify as a search directive---we obtained a diverse sample of 1.44M unique queries that varied widely in terms of both their content and structure. While not representative of what people are searching for today, these queries cover a wide range of topics (including music, sports, and politics), were produced across a 16 year span and include event-driven bursts (e.g., around the ICC Men’s T20 World Cup 2016, a biannual cricket tournament). 
These queries also widely varied in terms of their length, with the average search directive query containing an average of 4.5 words, which is slightly longer than estimates of query length in the US, which find that 82\% of queries are 3 words or less~\citep{keyworddiscovery2020keyword}. 
The longest query in our dataset was 896 tokens long, and 234 (0.01\%) queries were only one character, often an emoji.
Notably, Google Search limits queries to 32 words, and that length is counted after processing by an unknown tokenizer. When a query is too long, Google adds a notice at the top of the search results which states: ``... (and any subsequent words) was ignored because we limit queries to 32 words.''
We provide the distribution of query lengths with and without truncation in Appendix~\ref{sec:appendix-descriptives-queries}, Figure~\ref{fig:query-length-distributions}.

\subsection{Search Engine Results Pages (SERPs)}
\label{sec:methods-serps}

We used open-source tools to collect our search results (Section~\ref{sec:methods-serps-collecting}), and an iterative approach to discovering and classifying Google's warning banners (Section~\ref{sec:methods-serps-banners}).
To evaluate the rate at which search directive queries produce warning banners and data voids, we used our set of 1.4M unique queries as the inputs for an approach known as the algorithm audit~\citep{sandvig2014auditing}, which typically involves collecting and examining the outputs of a black-box system based on some fixed set of inputs (\cite{bandy2021problematic}, \cite{metaxa2021auditing}, \cite{mustafaraj2020case}, \cite{vanhoof2022searching}).
In this case, the inputs are the search directive queries, the system is Google Search, and the outputs are the Search Engine Results Pages (SERPs) returned by Google. 

\subsubsection{Collecting and Parsing SERPs}
\label{sec:methods-serps-collecting}

For collecting the search results available for each query, we used WebSearcher~\citep{robertson2020websearcher}---an open source tool for collecting and parsing SERPs that has been used in prior algorithm audits of Google Search~\citep{mejova2022googling}---to conduct a search using each query in our set, store the corresponding HTML, and extract details about its corresponding search results (e.g. rank, URL, result type). 
We also extracted several elements other elements from the SERP, including Google's estimate for the total number of results it found for each query (across its entire index), which could also be indicative of a data void, as the search results for a query with few matches may be easier to manipulate due to the limited competition.
As with most algorithm audits of web search, this SERP dataset represents only what someone searching these queries might have seen at the time of our collection. 
We also searched from a fixed location and do not study localization effects~\citep{kliman-silver2015location}.
Details on the number of results we collected are available in Appendix~\ref{tab:crawl-counts}

\subsubsection{Identifying Warning Banners}
\label{sec:methods-serps-banners}

We initially identified banners by checking for the exact phrasing of each warning banner type, and then built phrase-agnostic HTML parsers to extract them across the entire dataset. 
For low-relevance banners, the phrasing was ``It looks like there aren't many/any great matches for your search''~\citep{tucker2020getting}. 
Low-quality banners contained similarly phrased language (``It looks like there aren't many great results for this search''), swapping only ``matches'' with ``results.'' 
In contrast with the low-relevance banners, we never observed the variation where ``many'' was replaced with ``any'' in the low-quality banners, which aligns with the ambiguity of the banner message (``some of [these results] may not have reliable information'', Figure~\ref{fig:banner_ex}), and how they were described in Google's blog post announcing their rollout (``This doesn't mean that no helpful information is available, or that a particular result is low-quality''~\cite{nayak2022new}).
This reluctance to specify which search results are low-quality may also help explain why we never saw a low-quality banner for searches with a ``site:'' operator that restricted the search results to a specific web domain: doing so would remove the ambiguity of the judgment.
In contrast to these warnings about content, the rapidly-changing banner stated: ``It looks like the results below are changing quickly''~\citep{sullivan2021new}.
Additional details and a screenshot of the rapidly-changing banner, as well as details and a screenshot of a low-relevance banner variant that only appeared in our last crawl, are available in Appendix~\ref{sec:appendix-banners}.

\subsection{Search Query Text Features} 
\label{sec:methods-queries}

To evaluate query content, we used a dictionary-based approach to identify queries containing partisan and polarizing search terms or conspiracy-related search terms (Section~\ref{sec:methods-queries-polcon}), and other text features, such as advanced query operators (Section~\ref{sec:methods-queries-operators}).

\subsubsection{Political and Conspiracy-related Lexicons} 
\label{sec:methods-queries-polcon}

To identify queries around controversial topics that could potentially lead to data voids, we used a dictionary-based approach to tag words and phrases associated with conspiracies and politics in prior work.
Specifically, we used:
\begin{inparaenum}[(1)] %
    \item \textcite{ballatore2015google}'s set of 96 conspiracy-related search queries,
    \item \textcite{mahl2021nasa}'s set of 44 conspiracy-related hashtags, and
    \item \textcite{urman2022where}'s set of 6 conspiracy-related search queries.
\end{inparaenum}
We also considered \textcite{haak2023qbias}'s set of QAnon-related search queries and autocomplete expansions, but the terms were too broad for our purposes.
For \textcite{mahl2021nasa}, we added non-hashtag versions of each item (e.g. ``\#vaccineskill'' becomes ``vaccines kill'') and excluded ``\#dew'', which refers to conspiracies around directed energy weapons but produces a high false positive rate due to the popularity of Mountain Dew, a soda brand.
Of the 14 conspiracy categories covered by this dictionary---including conspiracies about 9/11, chemtrails, and reptilians---we found at least one search directive query that mentioned each.
We also used two existing lexicons of polarized terms---one designed to capture ``polarized language''~\citep{simchon2022troll} and one designed to capture ``partisan cues''~\citep{hu2019auditing}---to classify search directive queries as politically related.
Combined, we used these lexicons to classify the full set of queries as related to politics (11.1\%), conspiracies (0.11\%), both (0.02\%), or neither (88.8\%).

\subsubsection{Advanced Query Operators} 
\label{sec:methods-queries-operators}

Advanced query operators allow searchers to specify additional constraints on their search results.
For example, adding ``site:dailycaller.com'' to a query (e.g. ``trump site:dailycaller.com'') will search for results containing that term (``trump'') only within that site (``dailycaller.com'').
When considering search directives as an attempt to exert indirect online influence, the use of these operators has strategic value in guiding people to specific content via a trusted search engine: searchers less familiar with these operators may not understand that their results have been filtered, and while some search engines (e.g., DuckDuckGo) display a message to informs users that such a filter is active, Google does not~\citep{robertson2023identifying}.
In total, 1.5\% of our 1.4M unique queries contained one of 11 advanced operators, and among those, the most common operator was ``site:'' (92.0\%), followed by ``inurl:'' (2.8\%), ``filetype:'' (1.9\%), ``intitle:'' (1.0\%), ``ext:'' (0.5\%), ``before:'' (0.5\%), ``source:'' (0.4\%), ``related:'' (0.3\%), ``allintitle:'' (0.3\%), ``after:'' (0.2\%), and ``allinurl:'' (0.1\%).
These queries varied widely in their content and complexity, some containing multiple operators and others containing only one.
For example, one query used the OR operator, parentheses, quotes, and 15 site operators: ``(mask | vaccine | "death count" | "case count") fraud and evidence election ( site:amac.us | site:townhall.com | site:heritage.org | site:thegatewaypundit.com | site:oann.com | site:scienceunderattack.com | site:conservativetribune.com |  site:thefederalist.com |site:greatamericandaily.com | \\ site:westernjournal.com | site:zerohedge.com | site:prageru.com | site:realclearpolitics.com | site:mercola.com | site:naturalnews.com ).''

\subsubsection{Query Language}
\label{sec:methods-queries-language}

As many queries are names, fragments, emojis, or are otherwise grammatically incorrect, determining the language of queries can be challenging, and some level of noise is inevitable. To get a general sense of query languages, we used the FastText library~\citep{joulin2016bag} to predict the most likely language for each query. Across all 1.4M queries, more than 1.2M were predicted to be in English. For the subset of 930K queries where the model returned a confidence of at least 0.5, 875K were predicted to be English. The second most common category in the high and low-confidence query sets was French, with 23K and 9K queries, respectively. Many of the queries that were classified as French appear to have been classified that way because they use French words or names in an otherwise English-speaking context. For example, of the 55 queries that contained the name ``De Blasio''---the former mayor of New York---FastText predicted that 24 were French, including  ``bill de blasio'' and ``bill de blasio drops groundhog video''. 
German was the third most popular language, and similar to the French classifications, many of the queries classified as German appeared be English queries associated with American politics like ``adolf hitler defund police'' and ``f\"{u}hrermccarthy''.

\subsection{Web Domain Features}
\label{sec:methods-domains}

To evaluate the average domain quality of the SERPs we collected, we extracted the second and top level domain names for each URL (e.g. https://cnn.com/politics $\rightarrow$ cnn.com) and merged them with domain metrics for quality and partisanship (Section~\ref{sec:methods-domains-quality}), as well as domain-level measures of web traffic and backlink counts (Section~\ref{sec:methods-domains-seo}).

\subsubsection{Measuring Domain Quality and Partisanship}
\label{sec:methods-domains-quality}

For domain quality, we used a set of scores that were recently developed to evaluate the quality of domains based on a compendium of similar existing metrics~\citep{lin2023high}.
These scores range from 0 to 1, with higher scores indicating higher quality, and cover 11,519 unique domains (we drop one duplicate that appears with and without a ``www.'' prefix).
When calculating domain quality at the SERP-level, we take the average score of the domains that appeared on the SERP.
Prior to calculating that average, we exclude three platform domains from the original set because their quality scores were low and hard to interpret.
Those domains and their scores were: \nolinkurl{youtube.com} (0.375), \nolinkurl{facebook.com} (0.407), and \nolinkurl{google.com} (0.668). 
For partisanship, we use the partisan news scores created in~\citep{robertson2018auditing} based on the relative proportion of Democrats and Republicans that shared a domain on Twitter.
These scores range from -1 (only shared by Democrats) to 1 (only shared by Republicans), with a score of 0 meaning only that a domain was shared by an equal number of Democrats and Republicans (i.e., not ``neutral''), and we used the rank-weighted average of these scores for each SERP.
Our domain-level measures of quality are coarse-grained and do not account for instances, for example, where unreliable domains publish accurate webpages, or vice versa.
The partisan scores we used are subject to similar limitations~\citep{green2025curation}.

\subsubsection{Search Engine Optimization (SEO) Metrics}
\label{sec:methods-domains-seo}

Given the relevance of Search Engine Optimization (SEO), a billion dollar industry aimed at improving websites' search rankings, to questions about search results, we also obtained SEO features (e.g. backlink counts) and traffic estimates from Ahrefs ({\nolinkurl{https://ahrefs.com}), a large SEO company.
Recent work using data from Ahrefs has shown that some of its features are predictive of misinformation, and suggests that its traffic estimates are reliable (\cite{carragher2024detection}, \cite{carraghermisinformation}, \cite{williams2023search}). 
We provide additional details on the SEO features we used, including their validity, use in past work, and descriptive characteristics, in Appendix~\ref{sec:appendix-descriptives-seo}.

\subsection{Logistic Regressions}
\label{sec:methods-logit}

In our logistic regression models, the presence of each banner type (low-relevance or low-quality, separately) was our dependent variable, and our independent variables included factors both related to the query and the SERP it produced.
We used the \texttt{statsmodels} library in Python to fit our logit models with L1 regularization. More specifically, we set the regularization parameter (alpha) to 0.1, and used the L-BFGS algorithm as the solver with a maximum of 10,000 iterations and convergence and zero tolerances set to $1e-8$. 
We trained separate models for each dependent variable and crawl because the rules governing their appearance could change over time.

Our models for predicting low-relevance banners demonstrated moderate fit, with pseudo-$R^2$ values of 0.39, 0.15, 0.38 for crawls 1, 2, and 3, respectively (Appendix~\ref{sec:appendix-logit-low-relevance}, Tables~\ref{tab:logit-low-relevance-crawl1},~\ref{tab:logit-low-relevance-crawl2},~\ref{tab:logit-low-relevance-crawl3}). In contrast, and likely due to the smaller sample size, our models for predicting low-quality banners demonstrated lower fit, with pseudo-$R^2$ values of 0.13 for both crawls 1 and 2 (Appendix~\ref{sec:appendix-logit-low-quality}, Tables~\ref{tab:logit-low-quality-crawl1} \& Table~\ref{tab:logit-low-quality-crawl2}).

\subsection{Banner Stability and Consistency}
\label{sec:methods-stability}

\subsubsection{Rapid Data Collection}
\label{sec:methods-stability-temporal}

To better understand the relationship between SERP results and quality banners, we collected SERPs for the 301 queries that produced a low-quality banner in crawl-1 approximately every four hours from June 7, 2024 to June 24, 2024. 
We dropped data from two collection intervals due to technical issues, and dropped five queries that did not return search results at any time step (truncated to four words: children's clarity about search engine..., international intelligence "search manipulation..., "the virtuebios and mortality resolution", "miembro del instituto de investigación..., why face masks don’t work...). 
This left us with 73 SERPs for each of the remaining 296 queries that returned a quality banner in our initial collection. 
The two gaps in this collection lasted for 45 hours, between June 11 and June 13, and for 15 hours, between June 17 and 18.
We find similar results in a pilot version of this dataset that we collected in March 2024 over 34 time steps (with about 1.5 hours between each) without any gaps (Appendix~\ref{sec:appendix-consistency-pilot}).

\subsubsection{URL Similarity}
\label{sec:methods-stability-rbo}

To measure the stability of returned (ranked) URLs given a query, we use Ranked-Biased Overlap (RBO)~\citep{webber2010similarity},
a metric designed to compare ranked indeterminate lists. While RBO can be weighted by a user-chosen probability $p$, we elect to take the average overlap, which corresponds to RBO with $p=1$. RBO calculates the agreement between ranked lists $S$ and $U$ at every level of depth from $1:D$ and takes the average. Formally, RBO with $p=1$ can be written as:

\begin{equation}
    RBO(S,T, p=1) = \frac{1}{D} \sum^D_{d=1} \frac{\mid S_{1:d} \cap U_{1:d} \mid}{d}
\label{eq:AO}
\end{equation}

As our interest is in the relative stability of the SERPs across all queries, we constructed a windowed RBO metric ($RBO_k$) to quantify SERP similarity across consecutive pulls. 
Let $X_i$ be the pairwise RBO similarity matrix for URLs returned in the SERPs of query $i$ over $T$ timesteps. We set a window size $K$, and define query-level windowed RBO similarity $\bar{x}_i$ as:

\begin{equation}
    \bar{x}_{i,K} = \frac{1}{2KT} \sum^T_{t=0} \sum^K_{k=1} (\mathbbm{1}_{[t-k \geq 0]} x_{t-k} + \mathbbm{1}_{[t+k \leq T]}x_{t+k})
\label{eq:rbosim}
\end{equation}

\noindent
where $\mathbbm{1}$ is the indicator function. If we set $k=1$, this would correspond to the average RBO of URLs returned at time $t$ with URLs returned $t-1$ and $t+1$ over all time steps. If this number were close to 1, that would indicate a high similarity---both in the set of URLs returned and their ranking---between consecutive SERPs for a given query. Conversely, if $RBO_k$ were 0, this would suggest high volatility, and would mean that a query never returns any of the same URLs in consecutive time-steps. Finally, we define $RBO_k$ as:

\begin{equation}
    RBO_k = \frac{1}{N}\sum_{i=1}^N \bar{x}_{i,K}
\label{eq:rbok}
\end{equation}
\noindent
Intuitively, while $\bar{x}_{i,K}$ measures the stability of a single query's SERPs results over $K$ consecutive pulls, $RBO_k$ simply measures the average SERP stability over all queries' SERPs over $K$ consecutive pulls.

\subsubsection{URL Dependencies}
\label{sec:methods-stability-dependency}

To find a consistent and simple logic that could explain low-quality banner variance in all queries, we considered three formalized questions that probe simple URL dependency conditions. 
Specifically, we attempt to determine for all queries whether or not there is 1) a single URL in all bannered SERPs but no unbannered SERPs, 2) a pair of URLs that appears in all bannered SERPs but no unbannered SERPs and 3) a pair of URLs conditioned on a rank cut-off that appear in all bannered SERPs but no unbannered SERPs, e.g., ``If $u_i$ always appears in the top 5 search results and $u_j$ always appears in at a position below 5, is there always a banner?'' 
We provide a more formal treatment of these questions in Appendix \ref{sec:appendix-consistency-formal}.

\subsection{Model Development}
\label{sec:methods-models}

We consider models conditioned on several different features to attempt to learn a mapping between our queries and Google's low-quality banners. 
The purpose of these models is 1) to determine whether we can learn Google's mapping from observed queries and SERPs to quality banners and 2) to assess the consistency and stability of Google's approach to placing low-quality banners. 

\subsubsection{Preprocessing}
\label{sec:methods-models-preprocessing}
Prior to modeling, we performed two preprocessing operations. 
First, we calculated embeddings for all queries using a multilingual Sentence-BERT model \citep{reimers-2019-sentence-bert}. 
Second, most results on a SERP include a title---the blue text on Google's SERPs that one clicks to reach a webpage. 
For each domain that appeared at least twice in the dataset, we create a single string that contains the domain name with a colon followed by titles sampled from the domain with replacement. 
The intention with this step is to create a feature with some, albeit shallow, notion of the topics covered by the domain. 

\subsubsection{Train and Test Datasets}
\label{sec:methods-models-dataset}
After quantitative and qualitative evaluations, we chose a 1:3 positive-to-negative sample to train our classifiers.
In predictions on unlabeled data, we observed that the DistilBERT model trained on a 1:1 positive-negative sample seemed to be over-relying on the presence of quotation marks; DistilBERT's most 100 confident banner candidate predictions contained quotation marks and some contained names of movies or books like ``the craft'' and ``scout mindset.'' We therefore elected to use a 1:3 positive-to-negative sample in order to provide a more diverse set of negative samples. The sample consists of the 301 queries that produced a low-quality banner in crawl-1, and an additional 903 queries that were randomly sampled from the queries that did not receive such a banner. 
We applied a stratified 80/20/20 split to the final set of 1,204 labeled and unlabeled queries.

\subsubsection{Query-Only DistilBERT Model}
\label{sec:methods-models-distilbert}
First, we consider a model that only includes query text. 
This assumes that banner presence is independent of a query's returned SERP, which we know from temporally-dense observations is not the case. 
We include a text-only model to determine how much signal is present in text in a given time period and as baseline by which we can compare more complex models. 
More formally, for a given query $T$, let $p(B | T)$ be the probability of a quality banner appearing, conditioned on query text. 
Our first model assumes that the probability of a banner depends only on semantic cues present in the query text $T$. 
Specifically, we fine-tune a DistilBERT model~\citep{sanh2019distilbert} to predict $p(B | T)$ for each query. 
The model is trained for two epochs using only the raw query text using an Adam optimizer with a learning rate of $2\mathrm{e}{-5}$ and a linear warm-up scheduler.

\subsubsection{Query-SERP GNN Models}
\label{sec:methods-models-gnn}

Next, we sought a model that could incorporate the assumption that two different queries with highly similar SERPs should likely have the same banner status. 
We therefore elected to use Graph Neural Networks (GNNs), as this allows us to propagate domain-level context into query representations (e.g., as done with different features in~\citep{williams2024dredge}).   %
To do so, we constructed two simple homogeneous (GNN\textsubscript{\emph{Hom}}) and heterogeneous (GNN\textsubscript{\emph{Het}}) graph neural network models which incorporate the assumption that the presence of a banner $p(B)$ depends on both the query text ($T$) and the content associated with returned domains $S$. 
These models aim to predict the conditional probability $p(B | S, T)$, integrating information from both sources. 
We represent the problem as a bipartite query-to-domain graph, with one node set corresponding to queries and the other to domains. 
Our approach is also similar to the vaccine-related query-weblog graphs used in~\citep{chang2024measuring}, but we elect to leverage DistilBERT for query embedding as our query topics span a broader range of domains.

Given a set of queries $Q = \{q_1, q_2, \dots, q_n\}$ and a set of returned SERP domains $D = \{d_1, d_2, \dots, d_n\}$ we construct a homogeneous, bipartite graph $\mathcal{G}_{Hom} = (V,E)$ where domains and queries are treated as the same node types. Both node types have text-based node features, and labels $Y$ are a binary variable indicating the presence of a banner on $Q$. We additionally construct a heterogeneous graph, $\mathcal{G}_{Het} = (V, E)$ where where $V$ and $E$ are associated with a node type mapping function $\Psi : V \rightarrow A$ and an edge type mapping function $\Phi : E \rightarrow \phi$. In our setting, the set of node types are $A = \{Q, D\}$ and the set of edge types are $\Phi = \{domain-to-query, query-to-domain\}$.%

To incorporate the assumption that ranking changes in $top-k$ SERP results (with URLs held constant) should not alter banner presence, we do not weight edges in our networks. To allow information to propagate between queries, we exclude ``pendulum'' domains---those that only appeared once in our SERP data. For each of those domains we sampled at most 10 ``titles''---the blue text that appears on Google search results (generally the title of the article or webpage)---embedded the titles with DistilBERT, and took the simple mean. 
Although this is a relatively simple and coarse-grained approach that excludes many relevant domain-level signals, we demonstrate its effectiveness and leave the incorporation of more nuanced domain-level features to future work.

Both models consist of a GraphSage convolution with a dropout of 0.5 and ReLU activation, followed by a second GraphSage convolution and a final log-softmax activation function~\citep{hamilton2017inductive}. This results in a model with 526k parameters. GNN\textsubscript{\emph{Het}} uses a heterogeneous GraphSage convolution~\citep{Fey2019}. In this setting, where there are only two node types with bi-directional ties, this doubles the number of model parameters to 1.05M. We use an Adam optimizer with $\eta = 1\mathrm{e}{-3}$, and a weight decay of $5\mathrm{e}{-4}$, Cross Entropy Loss, and a Cosine Annealing Learning Rate Scheduler with $\eta_{min} = 2\mathrm{e}{-5}$~\citep{loshchilov2016sgdr}. 

\subsubsection{Model Validation}
\label{sec:methods-models-validation}
We evaluated our models using standard Accuracy, F1, Precision, and Recall metrics (Table~\ref{tbl:GNNResults}).
However, given the small size of the dataset and the corresponding likelihood of over-fitting, we also include three different supplemental forms of validation. 
For each of these supplemental evaluations, we consider unlabeled queries that each model most confidently predicted as unreliable. 
To evaluate the success of each model in identifying candidate queries for receiving low-quality banners, we used them to predict warning banners in the subsequent crawl of 1.4M SERPs and examined average SERP domain quality scores over the most confident predictions (Figure~\ref{fig:qualwindow}), evaluated annotated precision\@K and the prevalence of Children's Immortality Project queries in top results (Table~\ref{tbl:precisionCIP}), and evaluated a case study around a frequently observed set of search directives (Appendix~\ref{sec:appendix-cip}).

Some SERPs that returned results for crawl-1 did not return results for crawl-2 (8.8K) and vice versa (83K). Additionally, as with crawl-1, we only include domains which appeared at least twice across all of crawl-2 when running inference. To make results comparable across crawls, 
we use the intersection of queries on which we successfully ran inference in each crawl (1.4M). Figure~\ref{fig:qualwindow} contains only predictions on queries in this intersecting set, i.e., SERPs on which we could run inference in both crawl-1 and crawl-2.

\section{Acknowledgments}
The research for this paper was supported in part by the Stanford Internet Observatory, the Stanford Cyber Policy Center, the Office of Naval Research, MURI: Persuasion, Identity, \& Morality in Social-Cyber Environments under grant N000142112749 and by the Knight Foundation. It was also supported by the center for Informed Democracy and Social-cybersecurity  (IDeaS) and the center for Computational Analysis of Social and Organizational Systems (CASOS) at Carnegie Mellon University. The views and conclusions are those of the authors and should not be interpreted as representing the official policies, either expressed or implied, of the ONR or the US Government. 
Jeff Hancock was the faculty directory of this work at Stanford, and we are grateful to Renee DiResta, Stefan Feuerriegel, and Harry Yan for comments on earlier drafts.
The search data used in this study were collected using machines at Northeastern University that are administered by the authors' collaborators, with their permission. Northeastern University was given permission to query Google Search automatically for research purposes. Google did not review our research design, nor had any review rights with respect to the manuscript.
Prior versions of this paper were presented at the 2024 International Conference on Computational Social Science (IC2S2) and the 2024 International Workshop on Cyber Social Threats (CySoc).

\newpage

\phantomsection
\addcontentsline{toc}{section}{Appendix}
\section*{Appendix}
\label{sec:appendix}

\renewcommand{\thesubsection}{\Alph{subsection}}

\begin{appendices}

\subsection{Warning Banners}
\label{sec:appendix-banners}

Detailed counts and percentages (of all SERPs) for the warning banners we collected in each crawl are provided in Table~\ref{tab:crawl-banners}.
In the remainder of this section, we provide details and screenshots for Google's rapidly-changing banners (Appendix~\ref{sec:appendix-banners-rapidly}), as well as the low-relevance banner variants we observed across crawls (Appendix~\ref{sec:appendix-banners-relevance}). 

\begin{table}[hb!]
    \centering
    \scriptsize
    \caption{Warning banner types observed by crawl for our 1.4M search queries.}
    \label{tab:crawl-banners}
    {\renewcommand{\arraystretch}{1}
    \begin{tabular}{rrrr}
\toprule
\textbf{Warning Banner Type} & \textbf{crawl-1} & \textbf{crawl-2} & \textbf{crawl-3} \\
\midrule
No banner & 1,423,474 (98.997\%) & 1,424,269 (99.052\%) & 1,419,295 (98.706\%) \\
Any banner & 14,424 (1.0031\%) & 13,629 (0.9478\%) & 18,603 (1.2938\%) \\
Low-relevance (all) & 14,121 (0.9821\%) & 13,406 (0.9323\%) & 18,593 (1.2931\%) \\
Low-relevance (not many great matches) & 14,062 (0.9780\%) & 13,348 (0.9283\%) & 12,468 (0.8671\%) \\
Low-relevance (not any great matches) & 59 (0.0041\%) & 58 (0.0040\%) & 44 (0.0031\%) \\
Low-relevance (no matches) & - & - & 6,081 (0.4229\%) \\
Low-quality & 301 (0.021\%) & 221 (0.0154\%) & - \\
Rapidly-changing & 2 (0.0001\%) & 2 (0.0001\%) & 10 (0.0007\%) \\
\bottomrule
\end{tabular}

    }
\end{table}

\clearpage
\newpage

\subsubsection{Rapidly-changing Banners}
\label{sec:appendix-banners-rapidly}

We provide a limited examination of Google's rapidly-changing banners (Figure~\ref{fig:freshness_banner}) in the main text because we only observed two instances in the first two waves and 10 instances in the third wave. 
These 14 queries were unique with no repeats across crawls.
The queries triggering these banners were often political, with 11 of the 14 contained either ``trump'' or ``biden'' in the query.
For crawl-1 (October 2023), the queries were: ``biden zionist'' and ``dr. reiner fuellmich''. 
For crawl-2 (March 2024), the queries were ``republicans are russian asset'' and ``trump word salad''.
In crawl-3 (Sep 2024), the queries were: ``trump people magazine 1998'', ```lifelong republican''', ``trump people 1998 quote'', ``trump republicans dumbest voters'', ``trump republican voters dumbest'', ``"trump will win election"'', ``donald trump people magazine quote 1998'', ``trump wanders away|trump wanders away'', ``biden with young girls'', and ``trump tell republicans dumbest voters''.
The time-sensitive nature of these warnings is exemplified by the query ``biden zionist,'' which we received a rapidly-changing banner when we conducted that search on October 17, 2023, shortly after the Hamas attack on Israel on October 7, but did not receive any banner when we conducted it in our second or third waves.
Although our approach provides a limited sampling of these banners, it did surface a few, and future work could use those as seed queries for various query expansion methods to create a broader set of queries that might also trigger this type of warning for further examination.

\begin{figure}[th!]
\centering
\includegraphics[width=0.7\textwidth]{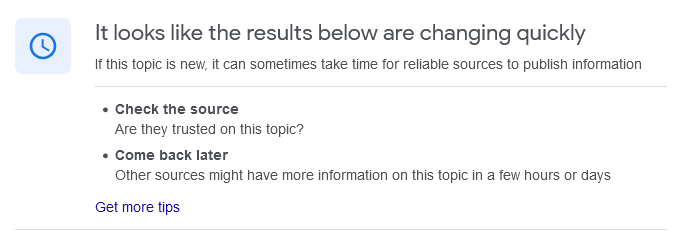}
\caption{Example of a freshness warning banner on Google Search. This banner is displayed when Google's systems detect ``a topic is rapidly evolving and a range of sources hasn't yet weighed in''~\citep{sullivan2010why}.}
\label{fig:freshness_banner}
\end{figure}

\subsubsection{Low-relevance Banner Variant}
\label{sec:appendix-banners-relevance}

In the third crawl, we observed a variant of the low-relevance banners that had not previously appeared (Figure~\ref{fig:topicality_banner_new}). 
This banner contained the same icon as the other low-relevance banners (a magnifying glass with a yellow background), but instead of stating that there were not many or any great matches, it states that ``Your search did not match any documents'' and provides a consistent list of generic suggestions for alternative searches (e.g., ``weather tomorrow'').
This variant appeared for 6,081 queries in crawl-3, accounting for 32.7\% of the 18,593 low-relevance banners we observed in total for that crawl. 
The presence of this variant was largely responsible for the overall increase in warning banner presence in that wave (from 1\% in crawl-1 to 0.9\% in crawl-2, to 1.3\% in crawl-3), despite low-quality banners disappearing (see Table~\ref{tab:crawl-banners}).
Although the exact language used in the banner is ``Your search did not match any documents'' (Figure~\ref{fig:topicality_banner_new}), we observed five cases in which the SERP contained a single generic ad (e.g., Dell, ADT). 

\begin{figure}[bh!]
\centering
\includegraphics[width=0.7\textwidth]{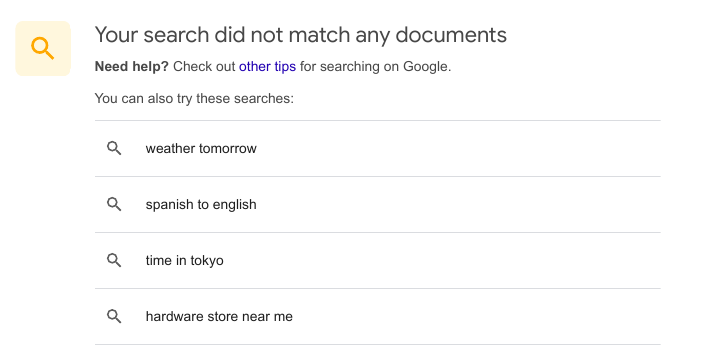}
\caption{Example of the low-relevance warning banner variant that we observed in crawl-3.}
\label{fig:topicality_banner_new}
\end{figure}

\clearpage
\newpage

\subsection{Descriptive Statistics}
\label{sec:appendix-descriptives}

Details on the data we collected in each crawl, including counts and averages for key measures, are provided in Table~\ref{tab:crawl-counts}.

\begin{table}[hb!]
    \centering
    \footnotesize
    \caption{Key metric counts and averages by crawl for our 1.4M search queries.}
    \label{tab:crawl-counts}
    {\renewcommand{\arraystretch}{1}
    \begin{tabular}{>{\raggedright\arraybackslash}rlll}
\toprule
& \textbf{crawl-1} & \textbf{crawl-2} & \textbf{crawl-3} \\
\midrule
Start Date & 2023-10-16 & 2024-03-10 & 2024-09-21 \\
Warning Banners (all types) & 14,424 & 13,629 & 18,603 \\
Search Results & 26,119,153 & 22,231,957 & 19,485,006 \\
Domains & 19,944,939 & 16,905,242 & 15,410,386 \\
News Domains & 2,391,776 & 1,810,744 & 1,463,805 \\
Unreliable Domains & 7,574 & 4,747 & 3,413 \\
Avg. Domain Quality & 0.783 & 0.788 & 0.792 \\
Avg. Domain Partisanship & -0.130 & -0.131 & -0.131 \\
Avg. Estimated Total Results & 280.8M & 289.9M & 204.4M \\
Avg. Domain SEO Traffic & 936.1M & 982.1M & 965.0M \\
\bottomrule
\end{tabular}

    }
\end{table}

\subsubsection{Search Queries}
\label{sec:appendix-descriptives-queries}

The length of the queries in our dataset (by token and character count) followed a heavy-tail distribution in its raw form (Figure~\ref{fig:query-length-distribution-full}) and remained skewed after truncating at Google's 32 token limit (Figure~\ref{fig:query-length-distribution-truncated}).

Without access to proprietary search engine data, it is impossible to know exactly how many users searched each query in our dataset. However, several large-scale query datasets have been released by search engines in the past. ORCAS, released by Microsoft, contains 10.4 million queries searched by at least ``$k$ different users, for a high value of $k$'' on Bing around January 2020 \citep{craswell2020orcas}. The exact value of $k$ used is not specified. The authors also applied filters to remove potentially offensive queries, like those containing pornography and hate speech. Although the ORCAS list is somewhat sanitized by its $k$-anonymity and offensive content filters, it allows us to check if any of our search directive queries were widely searched on Bing during that time period.

We find that 154,833 search directive queries were present in ORCAS, 1,972 of which were classified by our best-performing model as warranting a quality banner. These most confident 20 banner predictions include queries like ``facebook illuminati,'' ``lizard people conspiracy,'' and ``black groups that hate whites.'' The presence of these queries in ORCAS demonstrate both that data void queries exist on platforms beyond Google and that at least some of these queries are searched by a large number of users, by Microsoft's estimation. These queries could be spontaneous on the part of users, but given that the queries we analyze are all instances where one user publicly told at least one other user to search something---these queries meeting ORCAS $k$ threshold could be a result of the effectiveness of search directives.

\begin{figure*}[t!]
    \centering
    \begin{subfigure}[t]{0.98\textwidth}
    \includegraphics[width=\linewidth]{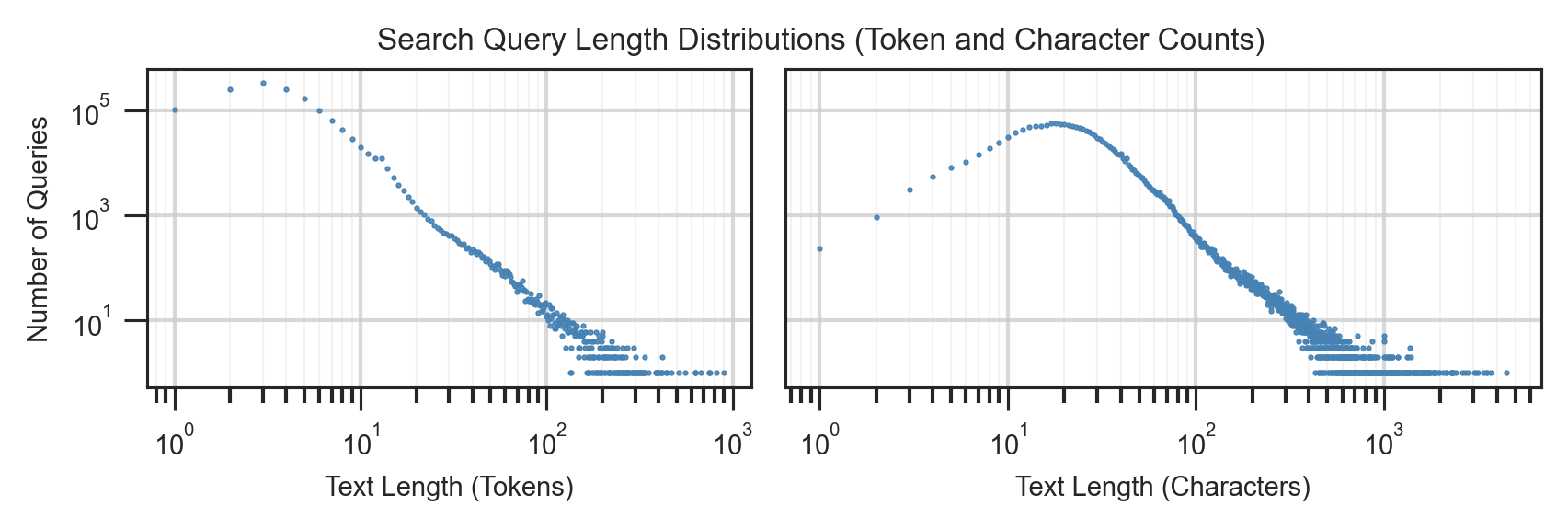}
    \caption{Full search query length distribution.}
    \label{fig:query-length-distribution-full}
    \end{subfigure}
    ~
    \begin{subfigure}[t]{0.98\textwidth}
    \includegraphics[width=\linewidth]{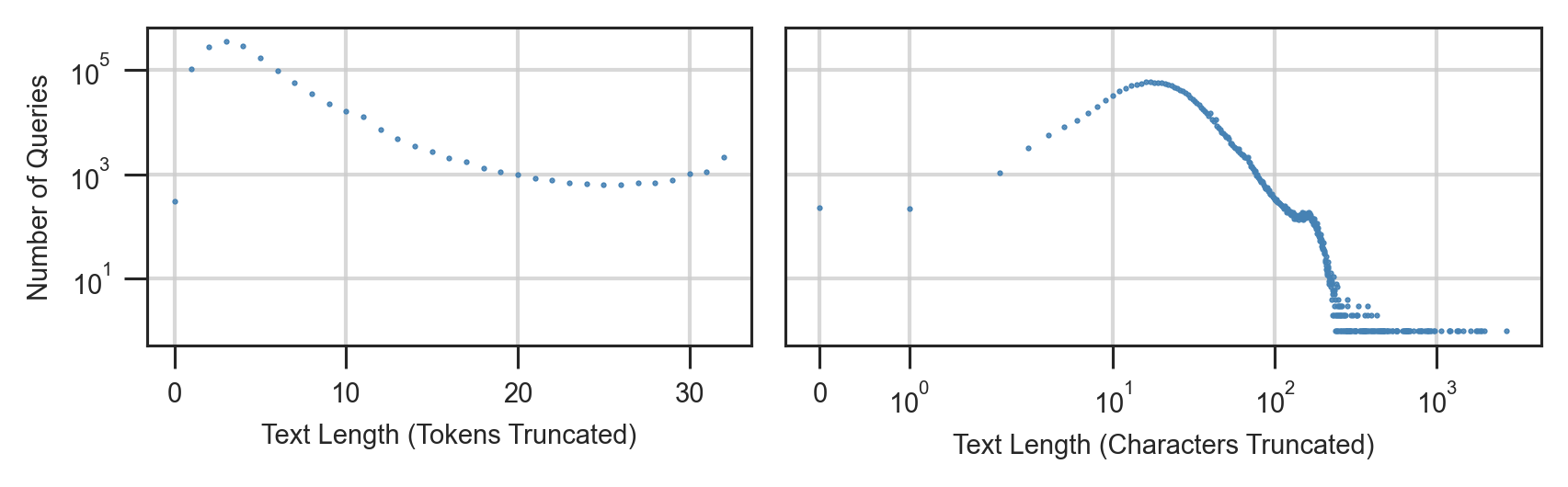}
    \caption{Truncated search query length distribution.}
    \label{fig:query-length-distribution-truncated}
    \end{subfigure}
    \caption{Search query length distributions for our set of 1.4M unique queries. The full distribution (a) shows the raw query lengths (measured via token and character counts), while the truncated distribution (b) shows the same distributions truncated at Google's 32 token query limit (see Methods~\ref{sec:methods-directives-queries}). Some queries have a truncated token count of 0 because they only contained punctuation or emojis (which were removed during tokenization), and a truncated character count of 0 because we measure that after truncating by tokens.}
    \label{fig:query-length-distributions}
\end{figure*}

\subsubsection{SERP Features}
\label{sec:appendix-descriptives-serps}
We found that Google Search's result size estimates follow a mixed distribution consisting of several heavy-tail distributions (Figure~\ref{fig:searches-length-distribution}).
While this distribution may indicate an underlying binning by Google, these estimates are generated through a non-public process, can vary based on a number of hard-to-control factors (like the data center used), and can vary in counter-intuitive ways. 
For example, longer and more specific queries (``cars -used'' are known to sometimes produce larger estimates than their shorter and more generic queries (``cars''), because the longer queries can cause the search engine to conduct a deeper search that surfaces a larger and more accurate estimate~\citep{sullivan2010why}.
While Google returned an estimate of 0 results for 1\% of queries in crawl-1, 0.87\% in crawl-2, and 1.3\% in crawl-3, around 50\% of those SERPs had one or more results (44.2\% in crawl-1, 43.9\% in crawl-2, and 66.5\% in crawl-3).

\begin{figure}[tb!]
    \centering
    \includegraphics[width=0.95\textwidth]{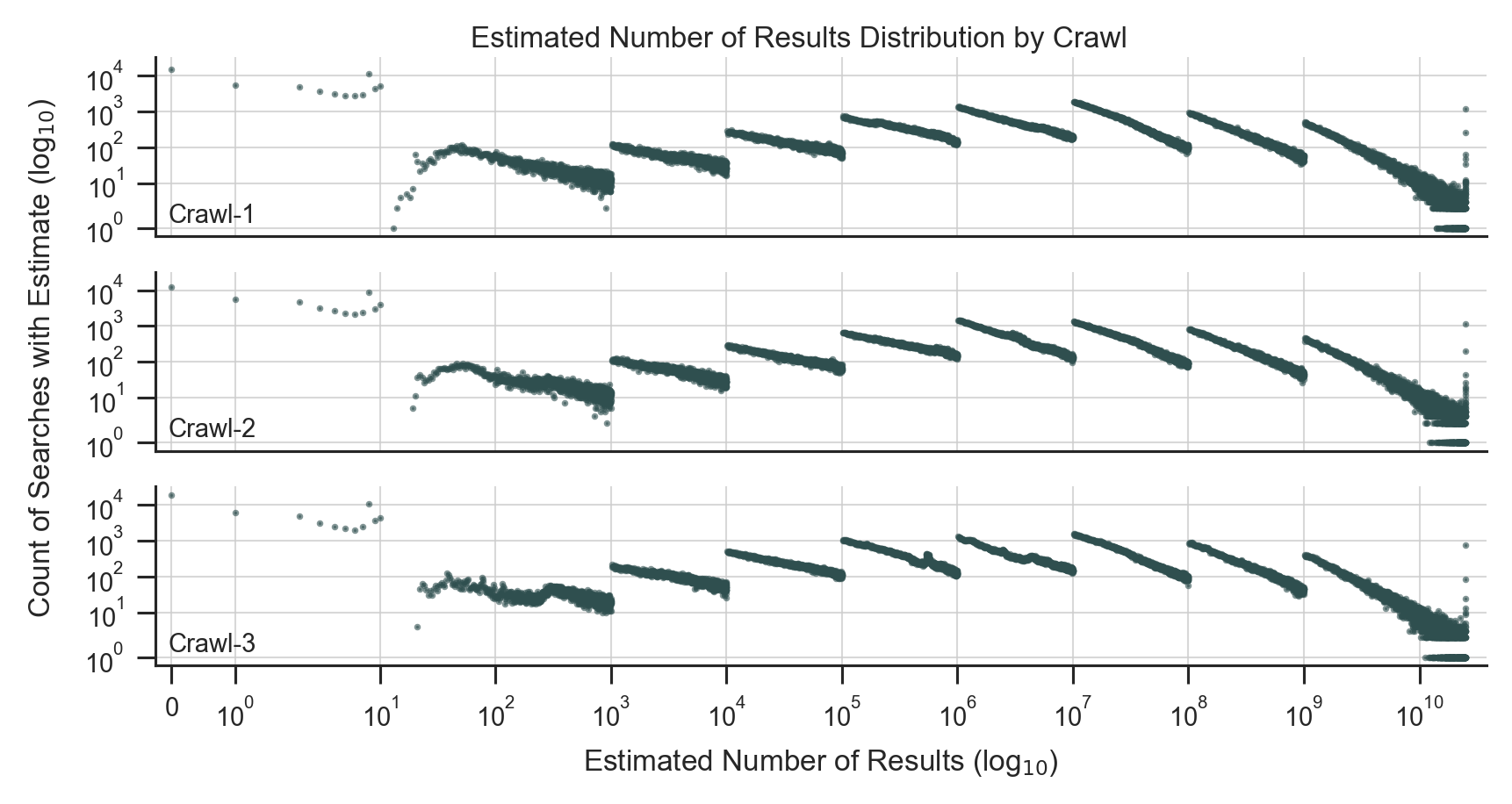}
    \caption{The estimated number of search results available for a given query---provided by Google in each SERP---follows a mixture distribution consisting of nine distinct distributions (e.g., 0 to 10 and 11 to 1000), suggesting different mechanisms for estimating the number of results within each band. The median number of estimated results was 4.6M in crawl-1, 3.3M in crawl-2, 1.9M in crawl-3, and 2.6M in crawl-4. Means ranged from 281M (crawl-1) to 204M (crawl-3) and standard deviations ranged from 1.4B (crawl-1) to 1B (crawl-4). Our data suggests a ceiling on Google's estimates, and the max value we observed in any crawl was 25.27B.}
    \label{fig:searches-length-distribution}
\end{figure}

\subsubsection{Domain Quality and Keyword Matches}
\label{sec:appendix-descriptives-quality}

Average domain quality was 0.78 (SD = 0.1), which is comparable to the score for \nolinkurl{aljazeera.com} (0.779) or \nolinkurl{tabletmag.com} (0.781).
In total, 159K queries (11.1\%) had a political keyword match, 1,588 (0.11\%) had a conspiracy-related keyword match, and 297 (0.02\%) had a keyword match for both.
Comparing average domain quality across these query types, we found small differences in their distributions when no banner was present, but no such differences when a banner was present (Figure~\ref{fig:search-quality-by-query-match}). 
When no banner was present, conspiracy-related queries produced SERPs with a higher average domain quality than SERPs produced by queries without such terms.
However, the overall differences in average domain quality by banner presence were much larger, and the average domain quality of SERPs that received a banner was about 10\% lower than average domain quality when no banner was present.

Among conspiracy-related queries, we found relatively small differences in the distribution of domain quality by specific conspiracy categories (Figure~\ref{fig:search-quality-by-conspiracy-category}).
The earliest linked search directive with a conspiracy-related query we found on Twitter occurred on July 19, 2009 and involved the longstanding conspiracy that 9/11 was an ``inside job'' (\cite{ballatore2015google}; \cite{mahl2021nasa}).
The post asked a question ``Was 9/11 an inside job?'' and provided a Google Search link with ``9/11 inside job proof'' as the query.
Given the leading wording of that query, it is possible that this would have led to a data void of conspiracy-related websites.
However, it is often difficult or impossible to reconstruct the search results one would have seen had they conducted that search in 2009.

\begin{figure*}[t!]
    \centering
    \begin{subfigure}[t]{0.48\textwidth}
    \includegraphics[width=\linewidth]{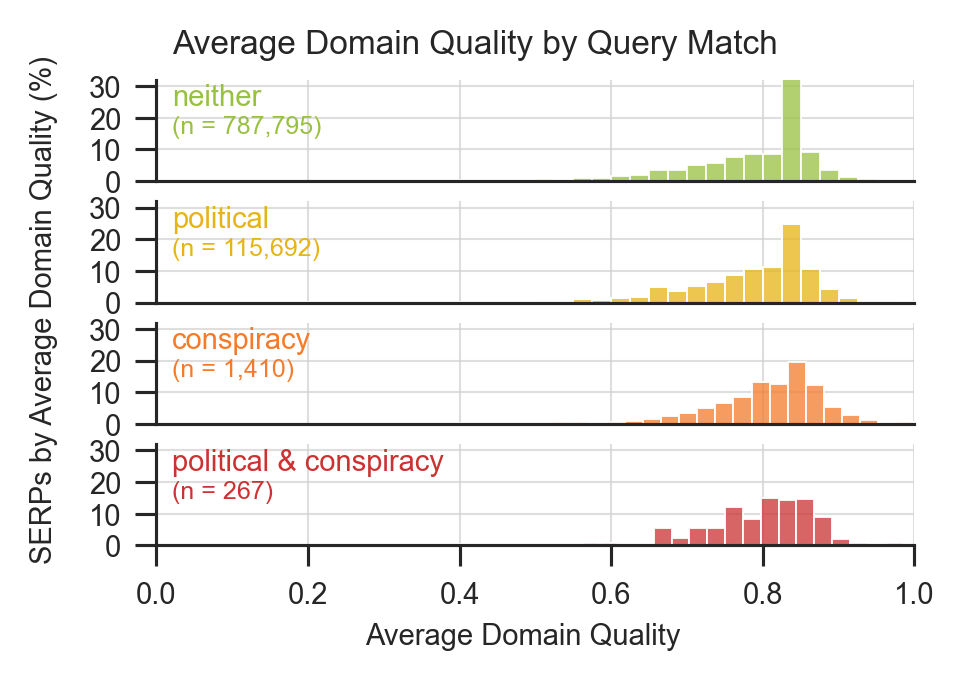}
    \caption{Distribution of average domain quality by query keyword matching. ``Neither'' indicates queries with no matches on the political or conspiracy lexicons, ``Political'' and ``Conspiracy'' indicate queries that match those respective lists, and ``Political \& Conspiracy'' indicates queries with matches from both.}
    \label{fig:search-quality-by-query-match}
    \end{subfigure}
    \hfill
    \begin{subfigure}[t]{0.50\textwidth}
    \includegraphics[width=\linewidth]{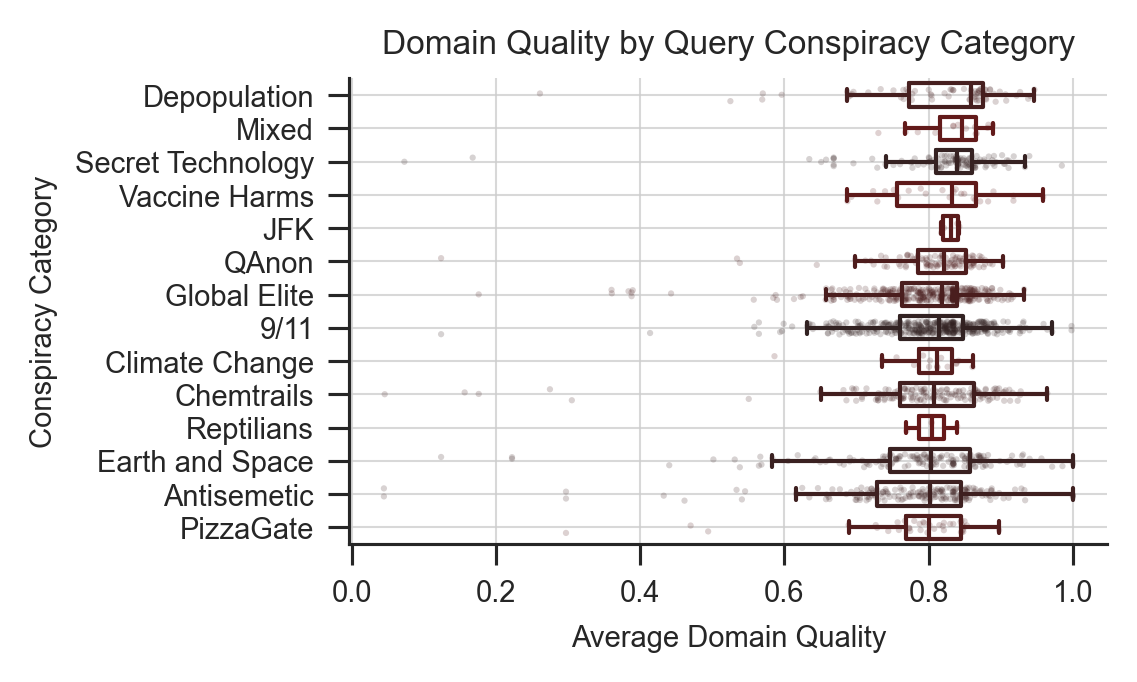}
    \caption{Average domain quality for conspiracy-related queries by conspiracy category. We did not find evidence of substantial variation in the average domain quality among the categories we examined. ``Mixed'' denotes queries that contained a mention of at least two unique categories.}
    \label{fig:search-quality-by-conspiracy-category}
    \end{subfigure}
    \caption{Domain quality by query type (a) and conspiracy category (b). There were few differences in domain quality by query type and conspiracy-related queries had the greatest average domain quality (0.80), above neither (0.78), political (0.79), and political \& conspiracy (0.79).}
    \label{fig:combined-search-quality}
\end{figure*}

\subsubsection{Search Engine Optimization (SEO) Metrics}
\label{sec:appendix-descriptives-seo}
Many SEO indicators require external data to calculate, which makes these indicators infeasible for researchers to collect given limited compute and storage resources, e.g., the number of backlinks a target domain receives is the total number of links that point towards the target domain from all other websites on the internet. Consequently, organizations and researchers must rely on third-party SEO toolkits. We use data from Ahrefs\footnote{\url{https://ahrefs.com}}, which had the 5th most active commercial webcrawler in October 2023 by number of requests according to Cloudflare Radar\footnote{\url{https://radar.cloudflare.com/traffic/verified-bots}}. Using Ahrefs' API, we extract 23 SEO features and 12 traffic estimation features for domains that ranked highly for conspiratorial or polarized keyphrases. Due to time and budget constraints, we elected to only extract this data for domains that appeared in at least 10 different conspiratorial or conspiratorial queries. These features include total number of backlinks, traffic estimates, number of backlinks from government domains, number of backlinks from educational domains, the amount of user generated content on the target domain, the number of unique referring IP addresses, and the number of unique referring domains. 

While the only way to obtain true website traffic is through website owners, a non-peer-reviewed case study by AuthorityHacker, which used traffic data provided by 47 website owners, found Ahrefs' traffic estimates to be the most accurate of any SEO toolkit \citep{ahrefstraffic}. Ahrefs' traffic estimates are calculated using position-weighted click-through rate estimates for Google search volumes of all keyphrases for which a website appears in the top 100 Google search results. Ahrefs self-reported similar results across a larger set of 1,635 domains\footnote{\url{https://ahrefs.com/blog/traffic-estimations-accuracy/}}. While these traffic estimates may contain noise, they are, to our knowledge, the best estimates currently available and have been used in past work on SERP reliability (\cite{carragher2024detection}, \cite{carraghermisinformation}, \cite{williams2023search}).

\para{SEO Features in Political or Conspiracy Queries}
Exploring the subset ($n=1,984$) of the domains with domain quality scores among all domains that ranked at least 10 times for political or conspiratorial keyphrases ($n=9,125$), we find that most SEO features exhibited a slight positive correlation with domain quality. 
There are notable outliers, including social media and video sharing sites, such as YouTube and Facebook, which have low domain quality scores but nonetheless receive 2.7B and 3.2B billion links from educational domains, respectively.
However, we also found small negative correlations between domain quality and a domain's number of webpages ($r=-0.06$), internal links ($r=-0.03$), external links ($r=-0.06$), and canonical tags (a way of signifying duplicate content to web crawlers) ($r=-0.07$). 
We additionally observe that these metrics all positively correlate with traffic. 

\begin{figure*}[t!]
    \centering
    \begin{subfigure}[t]{0.48\textwidth}
    \includegraphics[width=\linewidth]{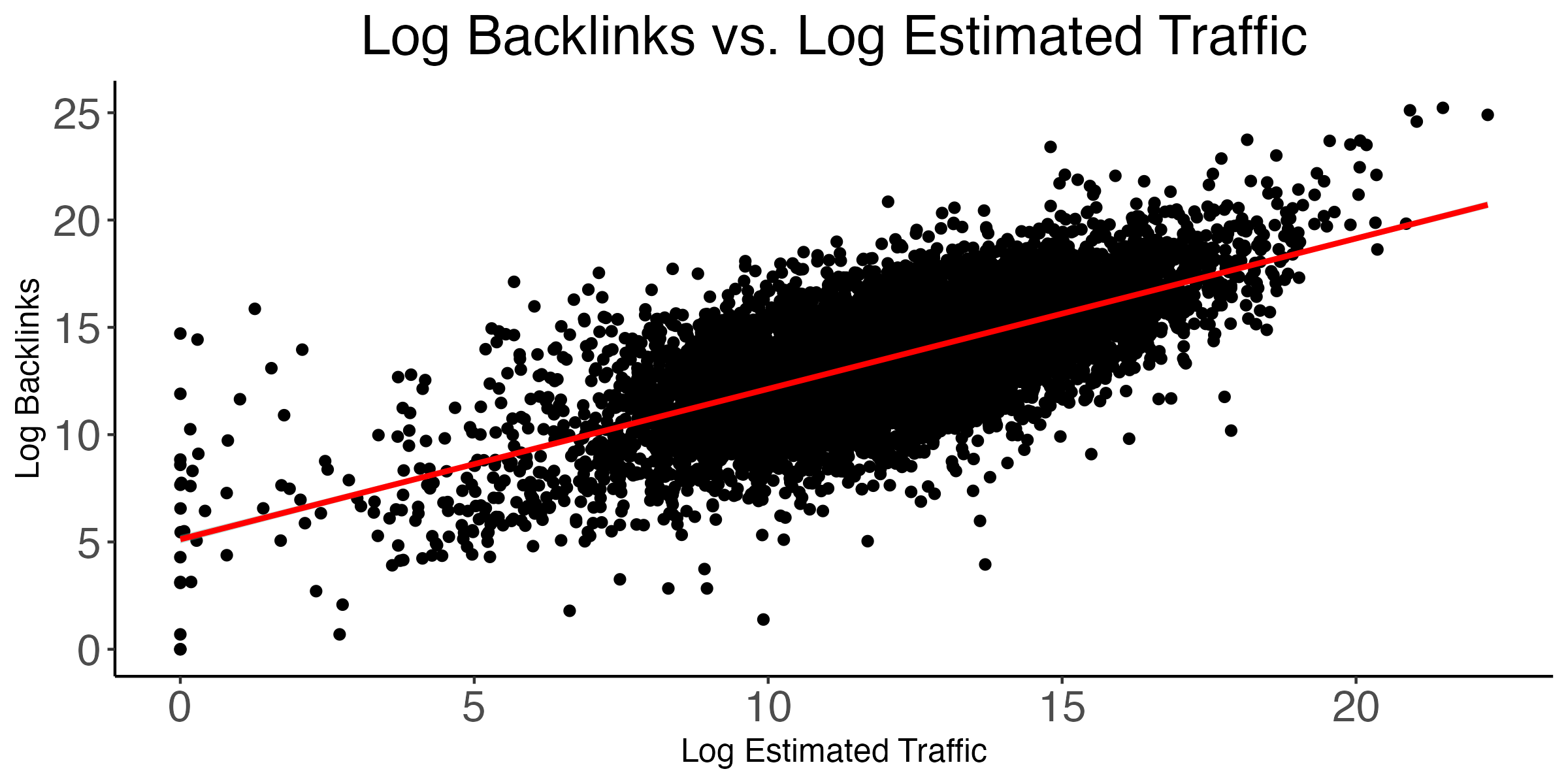}
    \caption{Log backlinks vs log estimated traffic over all conspiratorial or political domains that ranked for at least 10 queries.}
    \label{fig:seo_bl_tot}
    \end{subfigure}
    \hfill
    \begin{subfigure}[t]{0.49\textwidth}
    \includegraphics[width=\linewidth]{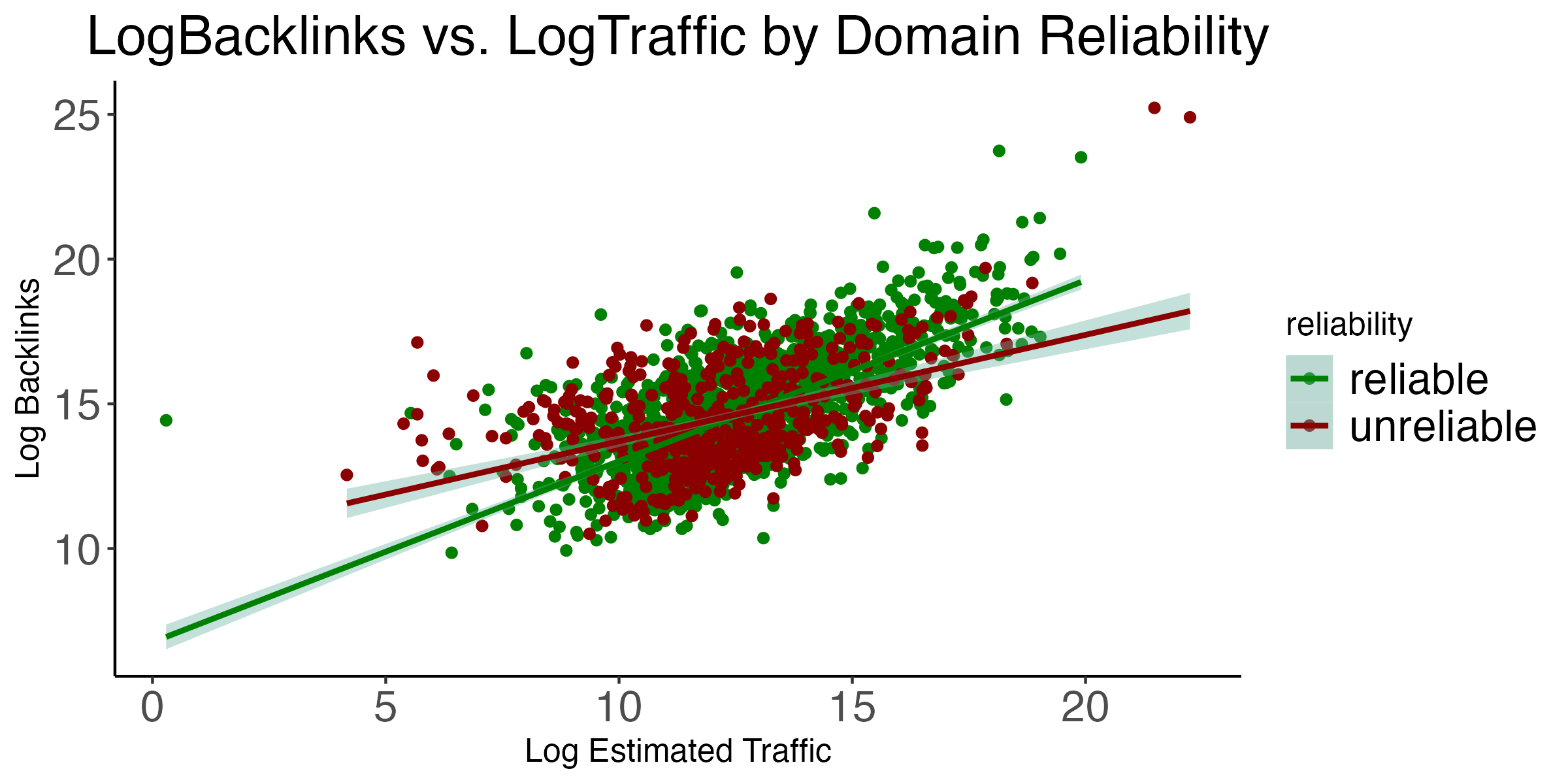}
    \caption{Log backlinks vs log estimated traffic for labeled reliable and unreliable domains.}
    \label{fig:seo_bl_rel}
    \end{subfigure}
    \caption{Number of backlinks compared to traffic estimates for domains that ranked for at least 10 political or conspiracy keyphrases (left) and for those which also have reliability labels (right). A 1\% increase in backlinks is associated with an approximate 0.77\% increase in traffic for reliable domains, and an approximate 0.67\% increase in traffic for unreliable domains.}
    \label{fig:seo_bl_traffic}
\end{figure*}

For the 9,125 domains that ranked at least 10 times for political or conspiratorial keyphrases, we observe a strong positive correlation ($r=0.78$) between backlinks and estimated traffic (Figure~\ref{fig:seo_bl_tot}). 
We partition the subset of these domains with quality scores ($n = 1,984$) by assigning the lowest quartile of the scores in our dataset (Domain Quality $< 0.62$) a label of ``unreliable'' and the remaining data a label of ``reliable.'' 
Using a simple interactive log-log regression model (where a log transformation is applied to the dependent and independent variables), we estimate that a 0.67\% change in traffic for 1\% change in backlinks for unreliable domains ($p<0.01$) and a 0.77\% change in traffic for a 1\% increase in backlinks for reliable domains ($p<0.01$; Figure \ref{fig:seo_bl_rel}). 

\subsection{Logistic Regressions}
\label{sec:appendix-logit}

We used logistic regression models to examine the factors associated with the presence of low-relevance and low-quality banners across our main crawls. 
The presence of either banner was our dependent variable, and our independent variables included factors both related to the query and the SERP it produced.
Among the query features were: character count (log10), presence of political keywords, presence of conspiracy keywords, and presence of query operators. We only used query operators as a feature for our low-relevance banner models because they never produced a quality banner in our dataset.
Among the SERP features were: average low-quality score of domains, rank-weighted news partisanship, estimated total results (log10), average domain traffic (log10), news domain count, and unique domain count.

We used the statsmodels library in Python to fit our logit models with L1 regularization. More specifically, we set the regularization parameter (alpha) to 0.1, and used the L-BFGS algorithm as the solver with a maximum of 10,000 iterations and convergence and zero tolerances set to $1e-8$. We trained separate models for each dependent variable and crawl because the rules governing their appearance could change over time.

\subsubsection{Low-relevance Banners}
\label{sec:appendix-logit-low-relevance}

Our models for predicting low-relevance banners demonstrated moderate fit, with pseudo-$R^2$ values of 0.39, 0.15, 0.38 for crawls 1, 2, and 3, respectively (Tables~\ref{tab:logit-low-relevance-crawl1},~\ref{tab:logit-low-relevance-crawl2},~\ref{tab:logit-low-relevance-crawl3}). 

\begin{table}[hb!]
    \centering
    \footnotesize
    \caption{Logistic Regression Results for Low-relevance Banners (crawl-1).}
    \label{tab:logit-low-relevance-crawl1}
    {\renewcommand{\arraystretch}{1}
    \begin{tabular}{>{\raggedright\arraybackslash}p{7cm}rlrr}
\toprule
\textbf{Variable} & \textbf{Coef.} & \textbf{95\% CI} & \textbf{Z} & \textbf{P} \\
\midrule
Query - Character Count (Log10) & 1.527 & (1.465, 1.590) & 47.846 & 0.000 \\
Query - Conspiracy Keywords (Bool) & 1.427 & (0.849, 2.006) & 4.838 & 0.000 \\
Query - Political Keywords (Bool) & 0.610 & (0.505, 0.715) & 11.351 & 0.000 \\
SERP - Average Low-Quality Score & 0.448 & (-0.014, 0.910) & 1.900 & 0.057 \\
SERP - Unique Domain Count & 0.044 & (0.025, 0.064) & 4.446 & 0.000 \\
SERP - Average Domain Traffic (Log10) & -0.205 & (-0.226, -0.185) & -19.527 & 0.000 \\
SERP - Estimated Total Results (Log10) & -0.228 & (-0.239, -0.218) & -43.667 & 0.000 \\
SERP - News Domain Count & -0.317 & (-0.346, -0.288) & -21.674 & 0.000 \\
SERP - Rank-weighted News Partisanship & -0.580 & (-0.766, -0.395) & -6.130 & 0.000 \\
Query - Contains Operator & -4.245 & (-5.566, -2.924) & -6.299 & 0.000 \\
Constant & -4.475 & (-5.016, -3.935) & -16.236 & 0.000 \\
\bottomrule
\end{tabular}
    }
\end{table}

\begin{table}[htb!]
    \centering
    \footnotesize
    \caption{Logistic Regression Results for Low-relevance Banners (crawl-2).}
    \label{tab:logit-low-relevance-crawl2}
    {\renewcommand{\arraystretch}{1}
    \begin{tabular}{>{\raggedright\arraybackslash}p{7cm}rlrr}
\toprule
\textbf{Variable} & \textbf{Coef.} & \textbf{95\% CI} & \textbf{Z} & \textbf{P} \\
\midrule
Query - Conspiracy Keywords (Bool) & 1.426 & (1.065, 1.786) & 7.753 & 0.000 \\
Query - Character Count (Log10) & 1.175 & (1.125, 1.224) & 46.790 & 0.000 \\
Query - Political Keywords (Bool) & 0.975 & (0.907, 1.043) & 28.118 & 0.000 \\
SERP - Rank-weighted News Partisanship & 0.420 & (0.283, 0.558) & 5.992 & 0.000 \\
SERP - Unique Domain Count & 0.169 & (0.153, 0.186) & 20.283 & 0.000 \\
SERP - News Domain Count & 0.069 & (0.057, 0.080) & 11.820 & 0.000 \\
SERP - Average Domain Traffic (Log10) & -0.107 & (-0.122, -0.091) & -13.325 & 0.000 \\
SERP - Estimated Total Results (Log10) & -0.119 & (-0.127, -0.111) & -28.212 & 0.000 \\
SERP - Average Low-Quality Score & -0.965 & (-1.346, -0.583) & -4.955 & 0.000 \\
Query - Contains Operator & -2.649 & (-3.315, -1.982) & -7.788 & 0.000 \\
Constant & -7.075 & (-7.510, -6.640) & -31.876 & 0.000 \\
\bottomrule
\end{tabular}
    }
\end{table}

\begin{table}[htb!]
    \centering
    \footnotesize
    \caption{Logistic Regression Results for Low-relevance Banners (crawl-3).}
    {\renewcommand{\arraystretch}{1}
    \label{tab:logit-low-relevance-crawl3}
    \begin{tabular}{>{\raggedright\arraybackslash}p{7cm}rlrr}
\toprule
\textbf{Variable} & \textbf{Coef.} & \textbf{95\% CI} & \textbf{Z} & \textbf{P} \\
\midrule
Query - Character Count (Log10) & 1.513 & (1.446, 1.579) & 44.381 & 0.000 \\
Query - Conspiracy Keywords (Bool) & 0.833 & (0.221, 1.445) & 2.666 & 0.008 \\
SERP - Average Low-Quality Score & 0.571 & (0.057, 1.086) & 2.176 & 0.030 \\
SERP - Unique Domain Count & 0.379 & (0.350, 0.408) & 25.951 & 0.000 \\
Query - Political Keywords (Bool) & 0.151 & (0.039, 0.264) & 2.634 & 0.008 \\
SERP - Average Domain Traffic (Log10) & -0.171 & (-0.194, -0.148) & -14.663 & 0.000 \\
SERP - News Domain Count & -0.241 & (-0.272, -0.210) & -15.150 & 0.000 \\
SERP - Estimated Total Results (Log10) & -0.303 & (-0.316, -0.290) & -45.517 & 0.000 \\
SERP - Rank-weighted News Partisanship & -0.690 & (-0.880, -0.499) & -7.101 & 0.000 \\
Query - Contains Operator & -1.592 & (-2.371, -0.812) & -4.003 & 0.000 \\
Constant & -7.478 & (-8.076, -6.879) & -24.487 & 0.000 \\
\bottomrule
\end{tabular}
    }
\end{table}

\newpage
\clearpage

\subsubsection{Low-quality Banners}
\label{sec:appendix-logit-low-quality}

In contrast to the models for the low-relevance banners, and likely due to the smaller sample size, our models for predicting low-quality banners demonstrated lower fit, with pseudo-$R^2$ values of 0.13 for both crawls 1 and 2 (Tables~\ref{tab:logit-low-quality-crawl1} \& Table~\ref{tab:logit-low-quality-crawl2}).

\begin{table}[htb!]
    \centering
    \footnotesize
    \caption{Logistic Regression Results for Low-quality Banners (crawl-1)}
    \label{tab:logit-low-quality-crawl1}
    {\renewcommand{\arraystretch}{1}
    \begin{tabular}{>{\raggedright\arraybackslash}p{7cm}rlrr}
\toprule
\textbf{Variable} & \textbf{Coef.} & \textbf{95\% CI} & \textbf{Z} & \textbf{P} \\
\midrule
SERP - Average Low-Quality Score & 4.903 & (3.505, 6.302) & 6.872 & 0.000 \\
Query - Conspiracy Keywords (Bool) & 4.032 & (3.206, 4.858) & 9.567 & 0.000 \\
Query - Political Keywords (Bool) & 0.624 & (0.045, 1.203) & 2.112 & 0.035 \\
SERP - Unique Domain Count & -0.039 & (-0.126, 0.048) & -0.881 & 0.378 \\
SERP - Rank-weighted News Partisanship & -0.102 & (-0.849, 0.645) & -0.266 & 0.790 \\
Query - Character Count (Log10) & -0.130 & (-0.557, 0.297) & -0.597 & 0.550 \\
SERP - Estimated Total Results (Log10) & -0.153 & (-0.205, -0.101) & -5.813 & 0.000 \\
SERP - Average Domain Traffic (Log10) & -0.197 & (-0.284, -0.110) & -4.448 & 0.000 \\
SERP - News Domain Count & -0.488 & (-0.662, -0.314) & -5.507 & 0.000 \\
Constant & -2.531 & (-5.176, 0.114) & -1.876 & 0.061 \\
\bottomrule
\end{tabular}
    }
\end{table}

\begin{table}[htb!]
    \centering
    \footnotesize
    \caption{Logistic Regression Results for Low-quality Banners (crawl-2)}
    \label{tab:logit-low-quality-crawl2}
    {\renewcommand{\arraystretch}{1}
    \begin{tabular}{>{\raggedright\arraybackslash}p{7cm}rlrr}
\toprule
\textbf{Variable} & \textbf{Coef.} & \textbf{95\% CI} & \textbf{Z} & \textbf{P} \\
\midrule
SERP - Average Low-Quality Score & 6.343 & (4.680, 8.007) & 7.473 & 0.000 \\
Query - Conspiracy Keywords (Bool) & 3.570 & (2.416, 4.724) & 6.062 & 0.000 \\
Query - Character Count (Log10) & 0.098 & (-0.428, 0.624) & 0.366 & 0.714 \\
SERP - Unique Domain Count & -0.020 & (-0.151, 0.111) & -0.301 & 0.763 \\
SERP - Estimated Total Results (Log10) & -0.152 & (-0.223, -0.080) & -4.144 & 0.000 \\
Query - Political Keywords (Bool) & -0.184 & (-1.061, 0.694) & -0.410 & 0.682 \\
SERP - Average Domain Traffic (Log10) & -0.267 & (-0.373, -0.161) & -4.920 & 0.000 \\
SERP - News Domain Count & -0.429 & (-0.639, -0.220) & -4.023 & 0.000 \\
SERP - Rank-weighted News Partisanship & -0.984 & (-1.927, -0.041) & -2.045 & 0.041 \\
Constant & -3.187 & (-6.492, 0.119) & -1.889 & 0.059 \\
\bottomrule
\end{tabular}
    }
\end{table}

\newpage
\clearpage

\subsection{Low-Quality Banner Consistency}
\label{sec:appendix-consistency}

\subsubsection{Formal Consistency Definitions}
\label{sec:appendix-consistency-formal}

In the main text we show that no URL pair conditioned on a rank cut-off can fully explain low-quality banner presence for all queries (Section~\ref{sec:evaluating}; Methods~\ref{sec:methods-stability-dependency}).
Here we provide a formal definition of the measure we used.
Let $Q =\{q_0, q_1, \dots, q_D\}$ be the set of 90 queries that have returned between 1 and 72 banners. 
Let $S = \{S_1, S_2, \dots, S_N\}$ be the set of SERPs that return a low-quality banner for a query $q_j$ where $S_i$ contains a ranked list of URLS $\{u_1, u_2, \dots, u_D\}$ and $R$ be the set of SERPs that do not return a banner for query $q_j$. 
Let $S_{:c*} =\{ S_{1,:c*}, \dots, S_{N,:c*}\}$ contain the bannered SERP results of all URLs below rank $c$ (e.g., $c=3$ would correspond to the first two URLs that appear on a SERP) concatenated with an arbitrary string `X', i.e. $S_{i,:c*} = \{\text{CONCAT}(u_1, `X'), \text{CONCAT}(u_2, `X'), \dots, \text{CONCAT}(u_{c-1}, `X')\}$. 
Conversely, let $S_{c:} = S_{1,c:}, \dots S_{N,c:}$ where $S_{i,c:} = \{u_c, u_{c+1}, \dots, u_D\}$ without concatenation. We define $R_{:c*}$ and $R_{c:}$ equivalently over unbannered SERPs. 
Finally let $S^*_{i,c} = (u_j \in S_{i,:c*}, u_k \in S_{i,:c*}) \cup (u_j \in S_{i,:c*}, u_k \in S_{i,c:}) \cup (u_j \in S_{i,c:}, u_k \in S_{i,c:})$. 
Again, we define $R^*_{i,c}$ equivalently using $R_{:c*}$ and $R_{c:}$. 
Using this formulation, we consider three questions about Google's system for placing low-quality banners:
\begin{packed_enumerate}
    \item : $\forall q \in Q$ does there exist a $u$ such that $\exists u \in \left( \cap_{i=1}^N S_i \right) \setminus \cup R$?
    \item : $\forall q \in Q$ does there exist a $(u_j, u_k)$ such that $\exists (u_j, u_k) \in \left(\cap_{i=1}^N S_i \right) \setminus \cup R$?
    \item : $\forall q \in Q$ does there exist a $c$ such that $ \exists (u_j, u_k) \in \left( \cap_{i=1}^N S^*_{i,c} \right) \setminus \cup R^*_c$?
\end{packed_enumerate}

\subsubsection{Rapid Data Collection Pilot}
\label{sec:appendix-consistency-pilot}

For the focused dataset we collected in June 2024, which used the subset of queries that produced a low-quality banner crawl-1 on a more rapid schedule (once every four hours), we had two gaps in the 73 time steps due to data collection issues. 
However, we also collected a pilot version of this dataset in which we searched the same query subset over 34 steps (once every hour-and-a-half) between March 10 and March 12, 2024.
This dataset had no gaps in its collection and three queries were dropped for not returning search results in any time step.
On average, 4\% of queries that had a low-quality banner in one time step did not have that banner type in the next time step, which is slightly higher than the 3.2\% change we found in June 2024.

The banner distribution was more bimodal than for our main dataset, with 116 queries having a banner for all 34 time steps, and 115 not having a banner at any time step. 
The remaining 67 queries had between 1 and 33 banners. 
We found the minimum Jaccard similarity across the 34 time steps banners was 0.82 and the mean was 0.89, which are close to the 0.79 min and 0.88 mean we found in the larger June 2024 dataset (Figure~\ref{fig:jaccard_qry_appendix}).
We recalculated $RBO_k$ on this data using a window size of 12 and again find the same general trend: every group is monotonically decreasing (Figure~\ref{fig:rbogroups_appendix}). 
Similar to the June 2024 dataset, we found that the queries that consistently returned a banner in every time step had the most stable search results. 
However, the relative ordering of other groups was somewhat different: queries that returned 16--33 banners across all time steps had the least stable search results in March 2024, but queries that returned zero banners across all time steps were the least stable in June 2024.

\begin{figure}[tb!]
    \centering
    \begin{subfigure}[t]{0.49\textwidth}
        \centering
        \includegraphics[width=\textwidth,trim={0 0 3cm 3cm},clip]{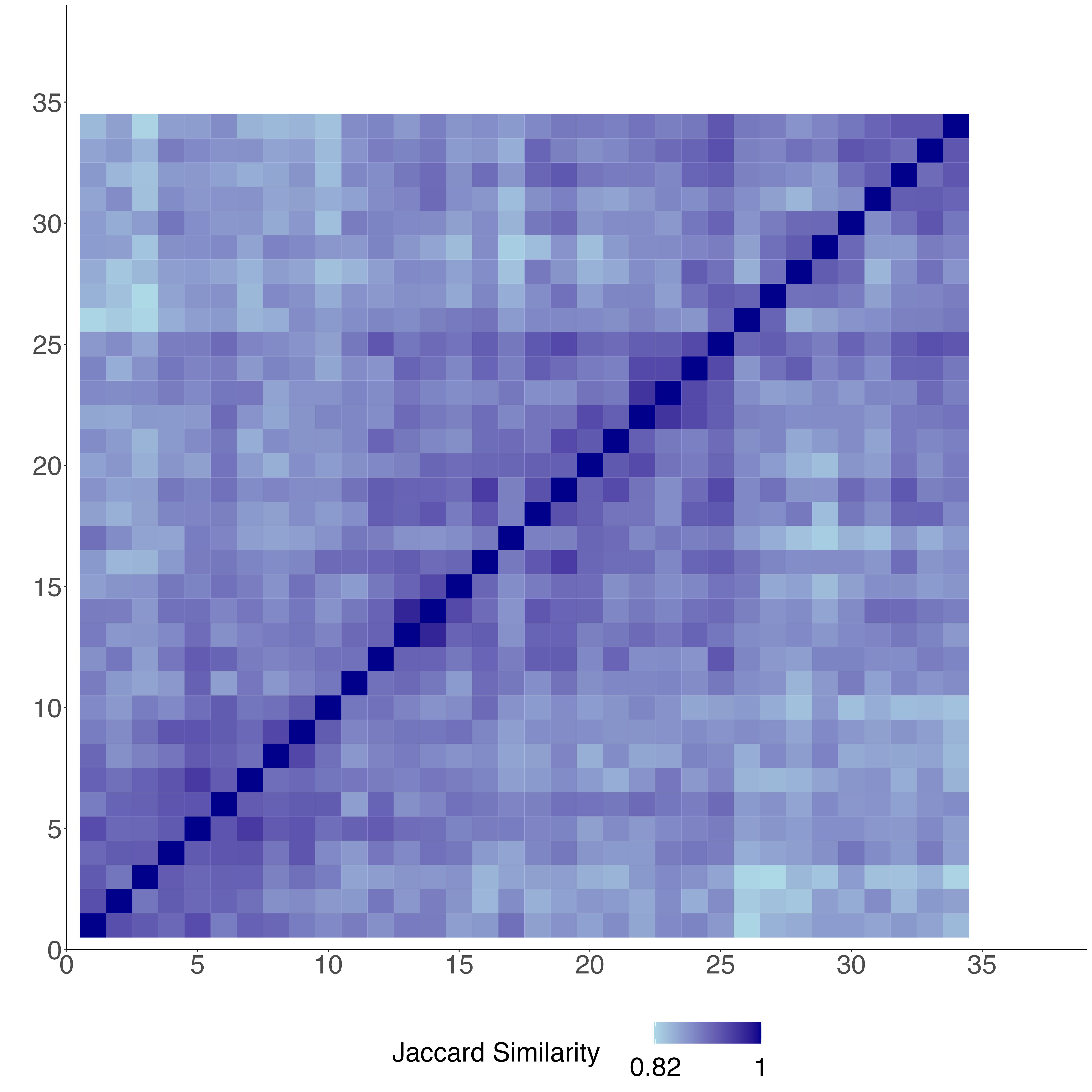}
        \caption{Heatmap cells show the pairwise Jaccard similarity of the set of queries that returned a quality banner over 34 time-steps.}
        \label{fig:jaccard_qry_appendix}
    \end{subfigure}
    \hfill
    \begin{subfigure}[t]{0.49\textwidth}
        \centering
        \includegraphics[width=\textwidth]{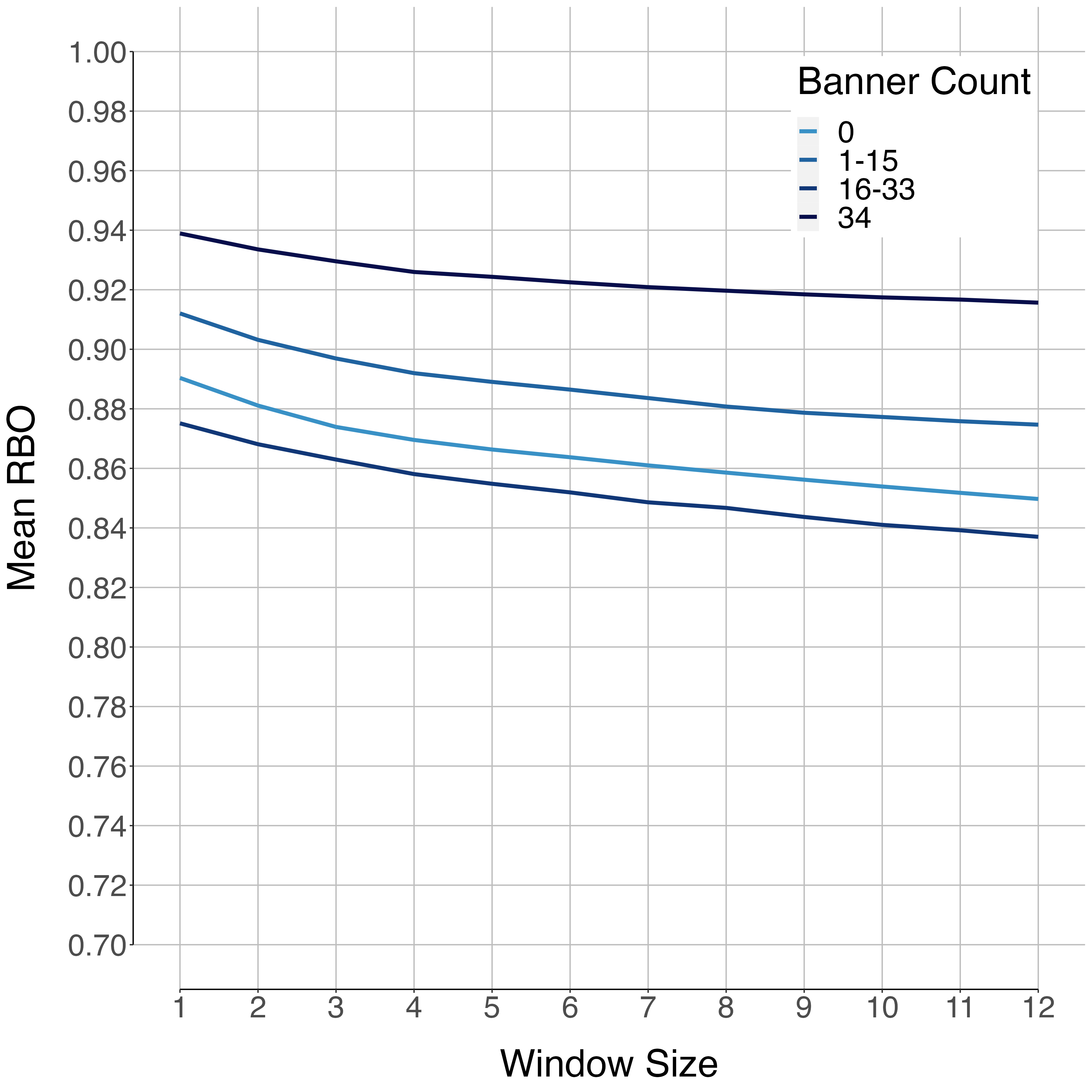}
        \caption{\(RBO_k\) (x-axis) over 12 window sizes (y-axis) by the mean RBO over all queries in the 34 time-steps.}
        \label{fig:rbogroups_appendix}
    \end{subfigure}
    \caption{Repeated Jaccard Similarity and RBO plots on 34 time steps with no collection gaps in March 2024 (Appendix~\ref{sec:appendix-consistency-pilot})}
\end{figure}

In both the 73 and 34 time-step datasets, we observe a bimodal distribution with the majority of queries having banners at all time steps or no time steps. These bimodal distributions make it unlikely that the instability in Google's low-quality banner placement could be the result of a stochastic process or of server latency issues. 
In the 73 timestep data, 166 queries never displayed low-quality banners and 40 always displayed banners. 
In the 34-timestep data, 115 queries never produced a low-quality banner, 116 always produced a low-quality banner, and 67 produced low-quality banners between 1 and 33 times. 
If either a stochastic process or server latency issues were responsible for inconsistencies in Google's low-quality banner placement, we would expect to see a more uniform distribution of banner instability over all time steps.

\subsubsection{Stability Experiments}
\label{sec:appendix-consistency-experiments}
We discussed the desirable property of ``stability'' in Methods and presented several examples of low-quality banner status changing as a result of minor query permutations (e.g., pluralizing or de-pluralizing nouns). In August 2024, we decided to perform several experiments to evaluate the stability of queries under various permutations. For example, in each of the 301 low-quality bannered queries, we pluralized all nouns and noun phrases, introduced keyboard typos with 10\% probability, and added or removed quotes for every query. To our surprise, not a single query generated a low-quality banner: only low-relevance banners remained. We initially assumed it may have been a geographic issue, so queries were repeated on computers in the east and west coasts of the US, and again, no low-quality banners were returned.

\subsection{GNN Models}
\label{sec:appendix-gnn}

\subsubsection{DistilBERT Model Results}

While the DistilBERT classifier yielded high accuracy on a small, balanced ``banner'' vs. ``non-banner'' query subset, we also found that it did not learn generalizable patterns from the query text data alone. When we ran inference over the 1.4M unlabeled queries using the fine-tuned DistilBERT model, we found its 100 most confident banner predictions all contained the presence of quotation marks, despite many of those queries not returning unreliable content (e.g., \textit{``phishing attack''}, \textit{``secretary of state''}, and \textit{``data breach''}). Reliance on such quotations is not ideal, because once a falsehood is subject to widespread fact-checking, a conspiratorial query can surface reliable or helpful results. In contrast, our graph neural network approach allowed us to condition the probability of a banner on both the text of the query and the reliability of its returned domains.

\subsubsection{Model Output Annotations}
Among the set of queries that produced a low-quality banner in October 2023, 141 appeared to be aimed at directing users to a network of strange websites aimed at boosting awareness of the ``Children's Immortality Project'' (CIP), which argues that children only die because adults teach them about death (see Appendix~\ref{sec:appendix-cip}). 
These queries often contained quoted keywords present in websites and were often surrounded by quotes (e.g., ``creating all children to die'', ``the virtue bios and mortality resolution'' and clarity genius of sedona ``interneted''). 
Given that these queries generated a large portion of the observed low-quality banners, the annotators also evaluated whether each query produced a SERP that surfaced websites related to the CIP.
We found substantial agreement between annotators on this task as well ($\kappa = 0.80$), with the homogeneous and heterogenous models producing similar results, but the homogeneous model identified more in its top 20 predictions (Table~\ref{tbl:precisionCIP}).

\begin{table}[t!]
\centering
\small
\caption{Precision of the discovery process for each model at the top 5, 10, and 20 predictions. We additionally report the number of Children's Immortality Project (CIP) queries returned in the top 20 predictions.}
\label{tbl:precisionCIP}
\begin{tabular}{lcccc}
\toprule
                              & P@5            & P@10           & P@20           & CIP Queries \\
\midrule
DistilBERT                    & 0              & 0              & 0.050          & 5           \\
GNN\textsubscript{\emph{Hom}} & 0.600          & 0.450          & 0.575          & \textbf{16} \\
GNN\textsubscript{\emph{Het}} & \textbf{0.800} & \textbf{0.900} & \textbf{0.775} & 15          \\
\bottomrule
\end{tabular}

\end{table}

\subsubsection{Out-of-Sample Banner Identification}
\label{sec:appendix-gnn-oos}

In crawl-2 there were 74 queries that received a low-quality banner that did not have a low-quality banner in crawl-1. 
To evaluate out-of-sample banner identification, we ran inference over all queries that did not receive a low-quality banner in crawl-1, and evaluated the number of those queries that received a low-quality banner at time step 2, sorted by confidence. 
Similar to our other model validation tests, we again find that the GNN models outperformed the DistilBERT model (Table~\ref{tbl:OOD_comparison}).

\begin{table}[t!]
    \centering
    \caption{The Homogeneous and Heterogeneous GNNs best identify new out-of-sample bannered queries. The number of the 74 new low-quality banner queries present in each model's $K$ most confident unreliable predictions.}
    \label{tbl:OOD_comparison}
    {\renewcommand{\arraystretch}{1.2}
    \begin{tabular}{lccccccc}
\toprule
Model                         & Top 10 & Top 50     & Top 100    & Top 500     & Top 1k      & Top 5k      & Top 10k \\
\midrule
DistilBERT                    & 0      & 1          & 2          & 9           & 12          & 19          & 21          \\
GNN\textsubscript{\emph{Hom}} & 0      & \textbf{3} & 5          & \textbf{13} & \textbf{16} & 24          & 27          \\
GNN\textsubscript{\emph{Het}} & 0      & 1          & \textbf{6} & 11          & 15          & \textbf{32} & \textbf{39} \\
\bottomrule
\end{tabular}

    }
\end{table}

\subsection{Query Examples}
\label{sec:appendix-examples}

\subsubsection{``Vril Lizards''}
\label{sec:appendix-examples-vril}

In a qualitative exploration of quality-bannered queries, we find that banner presence can also depend on relatively minor changes in a query. 
For example, we observed that the query \textit{vril lizards}, which refers to mythical parasitic lizards that some websites claim control the brains of celebrities, frequently returned a banner, but the query \textit{vril lizard} (where lizard is singular instead of plural) did not. 
However, when quotation marks were added to the singular query, ``\textit{vril lizard}'' Google did return a banner (Figure~\ref{fig:appendix-vril}, 2nd panel).
Although the quotes did not change the semantic content of the query, and the results for both queries contained multiple webpages that were either identical or exhibited nearly identical text content, only the quoted one received a banner. 
We additionally observed several other queries that only returned banners when quotation marks were placed around the query. 
Stranger still, when ``\textit{vril lizard}'' was searched the next day, its banner was gone. Together, these observations provide some evidence that Google's low-quality banners do not appear to depend solely on the semantic content of a query. 

\begin{figure}[tb!]
\centering
\includegraphics[width=\textwidth]{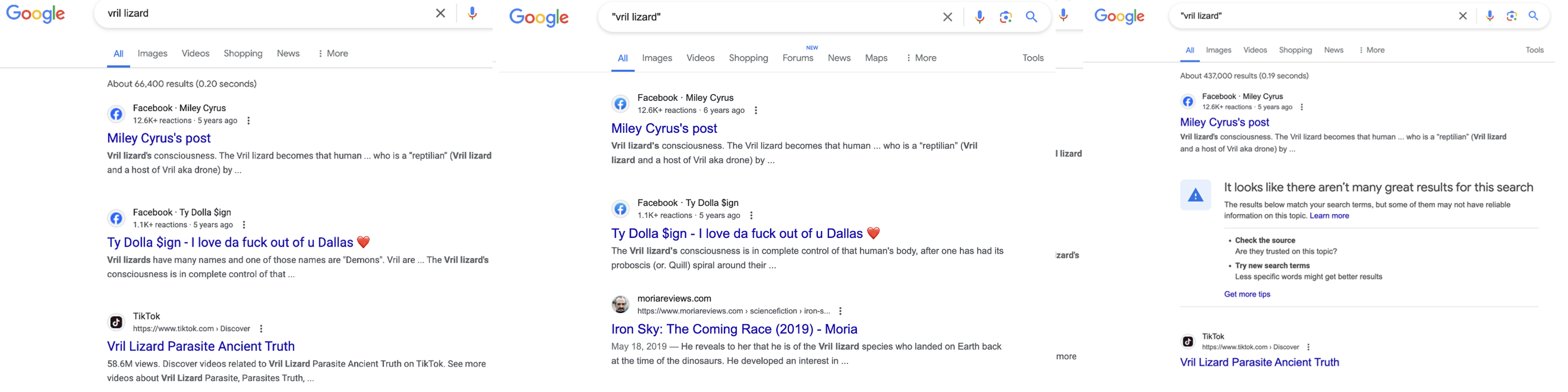}
\caption{Search results for the query \textit{vril lizard} without (left screenshot) and with quotation marks (right two screenshots). All SERPs contain a similar set of URLs. The quoted version received a banner in one instance, but not in second screenshot taken a day later.}
\label{fig:appendix-vril}
\end{figure}

\subsubsection{``MRNA Prions''}
\label{sec:appendix-examples-prions}

One search query from our dataset, ``MRNA prions,'' is likely in reference to debunked claims stemming from a 2021 report on how MRNA vaccines could cause diseases like Alzheimers~\citep{mrnaprions2021}. 
As of September 2024, the top 10 search results for ``MRNA Prions'' still surfaced support for this narrative, including a Research Gate article titled ``COVID-19 RNA Based Vaccines and the Risk of Prion Disease,'' the author of which has also been published by SciVision Publishers, which is included on a list of Predatory Journals and Publishers (\nolinkurl{beallslist.net}).
The ``MRNA prions'' query returned a low-quality banner in 56 of the 73 timesteps in our temporally dense dataset from June 2024, and we found that two URLs could explain the majority of instances (43 of 56) in which that query produced a banner using our rank-cutoff approach (Appendix~\ref{sec:appendix-consistency}).

\subsection{August 2024 Core Update}
\label{sec:appendix-aug2024}

The disappearance of quality banners coincided with an August 2024 Google core update \citep{aug2024update}. In this section, we consider the question ``did Google remove quality banners because the August 2024 update removed all unreliable results?'' We find that while the update did significantly impact search results for formerly quality-bannered queries SERPs did not necessarily become more reliable.

For each query, for each of the 73 time steps from June 2024, we calculated the Jaccard similarity and RBO between the top 10 SERPs of the June data and the August data (Appendix~\ref{sec:appendix-consistency-experiments}). We found that the mean average Jaccard similarity over all queries and time-steps was 0.31 (max 0.75), and the mean RBO was 0.37 (max 0.70). This was only slightly less than the SERPs with low-quality banners from March 2024 (Appendix~\ref{sec:appendix-consistency-pilot}), where the average Jaccard similarity and RBO with the June data were 0.27 (max 0.69) and 0.27 (max 0.62), respectively.

In our results section, we stated that 25 queries that displayed between 1 and 72 banners had at least one URL appear in all bannered SERPs that never appeared in unbannered SERPs. We looked for these URLs in the August data, and found that 8 queries displayed those URLs strongly associated with banners in the June data, yet returned no banner. Additionally, we calculated average domain reliability of the March 10th crawl, June 7th crawl, and August crawl. This is a coarse-grained approach, as we do not have labels for 81--90\% of domains returned in SERPs, but this still provides some frame of comparison. We found that the average domain-level SERP reliability (where 1 is most reliable, and missing values are ignored) were 0.52 (March), 0.44 (June), and 0.47 (August). While the labeled August SERPs results were, on average, slightly more reliable than the June 7th SERPs, they were slightly less reliable than the March SERPs. This provides additional evidence that banners were not turned off in August because the problem was solved.

A more qualitative exploration revealed that while many search results changed between June and August, the August search results often did not correspond with an increase in overall SERP reliability. The query ``mrna prions'' (Appendix \ref{sec:appendix-examples-prions}), did not display the newstarget or 4chan URLs strongly associated with a banner in Figure \ref{fig:mrnaq1}, but returned three URLS we never observed in the June data: two links to a linkedin post and youtube video of posted by an apparent Australian anti-vaccine organization\footnote{\url{https://web.archive.org/web/20240326082456/https://healthallianceaustralia.org/}} and a link to an 8kun post encouraging users not to get vaccinated\footnote{\url{https://web.archive.org/web/20240703140131/https://8kun.top/freedomzine/res/15841.html}}.

\begin{figure}[tbh!]
\centering
\includegraphics[width=0.5\textwidth]{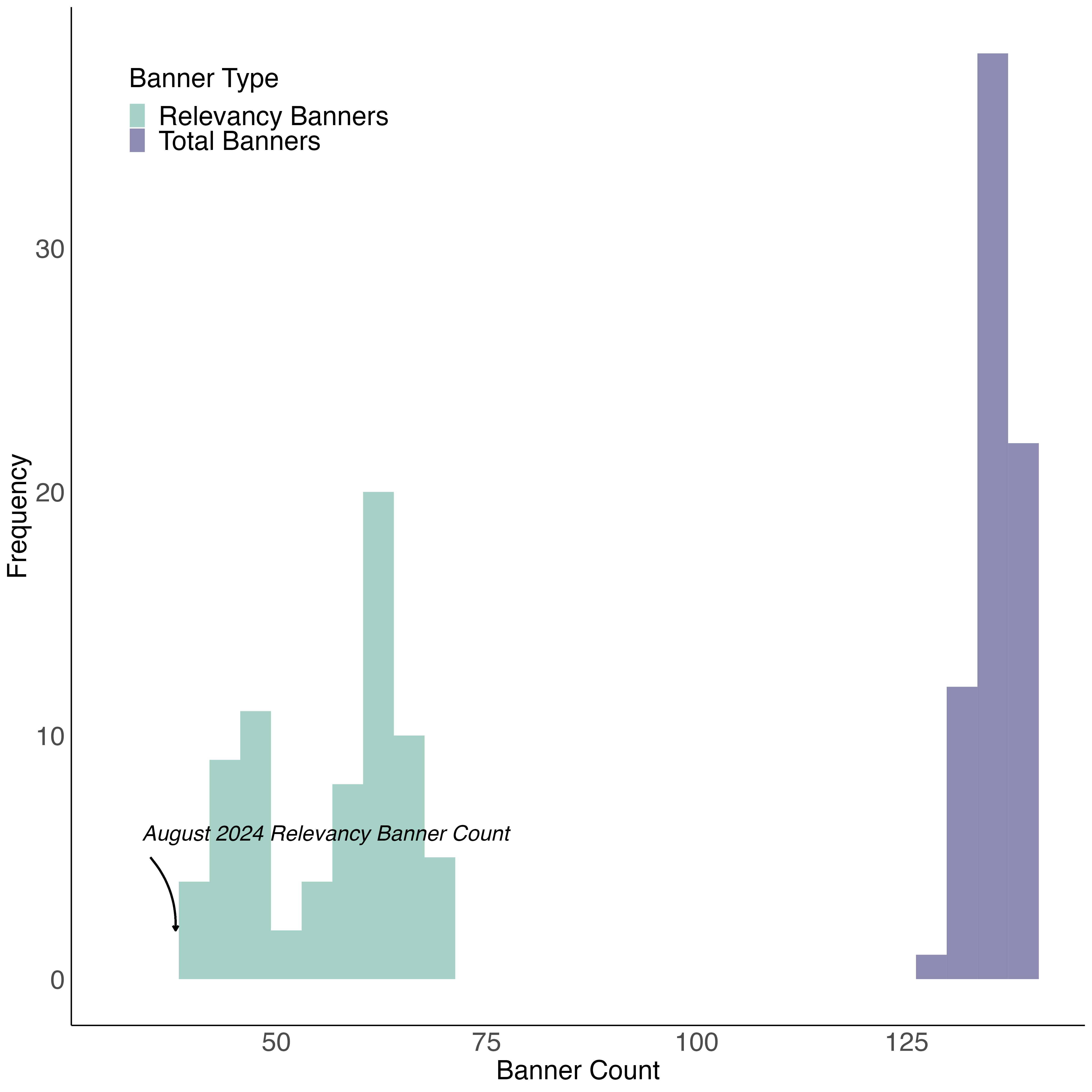}
\caption{Quality banners do not appear to have been subsumed by low-relevance banners. We display histograms of the distributions of the counts of low-relevance banners over 73 time-steps (teal) and the distribution of total banners (i.e., low-relevance + low-quality) (purple). In August 2024, the count of low-relevance banners we observed was lower than we observed across any of the 73 June 2024 time-steps.}
\label{fig:aug_relevance}
\end{figure}

We considered the possibility that quality banners had been subsumed inside low-relevance banners during the August 2024 update. 
To test this hypothesis, we created one histogram showing the total number of low-relevance banners for the 296 queries we analyzed over 73 time-steps from June 2024 and a second histogram containing the sum of the counts low-quality and low-relevance banners that appeared in each of those 73 time-steps (Figure~\ref{fig:aug_relevance}). 
If every quality banner were replaced with a low-relevance banners, we would expect that the total number of low-relevance banners in our August 2024 data should fall within or near the distribution of total banners (i.e., the sum of low-quality and low-relevance banners). 
This was not what we observed; rather, the number of low-relevance banners returned over the 296 queries in August (38) was lower than the count of low-relevance banners returned in any of the 73 time steps (the minimum was 41).

\subsection{Children's Immortality Project}
\label{sec:appendix-cip}

The self-proclaimed creator of the Children's Immortality Project (CIP) posted on many of his websites that Google was censoring him, and believed that by creating SEO websites, more people would learn of his message. 
We found hundreds of keyword-heavy SEO sites related to CIP that heavily linked to one another and often had questionable domain names (e.g, howtokillchildrenlegally-breed.blogspot.com\footnote{\url{https://web.archive.org/web/20240411055816/http://howtokillchildrenlegally-breed.blogspot.com/}}, keywordsarefortakingovertheinternet.blogspot.com\footnote{\url{https://web.archive.org/web/20240605114527/http://keywordsarefortakingovertheinternet.blogspot.com/}}, 
and google-bomb.com\footnote{\url{https://web.archive.org/web/20141217044958/http://google-bomb.com/}}), engendered discussion and speculation regarding possible intentions of these websites.\footnote{Explicit content: \url{https://web.archive.org/web/20100728210054/http://4chanarchive.org/brchive/dspl_thread.php5?thread_id=3595190\&x=Childrens+Immortality+Project}}

For each of the domains returned alongside a low-quality banner, we extracted the 10 domains which link most frequently to the target domain using Ahrefs. 
We used this data to create a bipartite network of second and top level domains to queries and ran Louvain community detection on the resulting graph.
One cluster stood out as highly interconnected and contained many of the domains amplifying the CIP (Figure \ref{fig:app_cip}). 
While we demonstrate that the backlink network displays some signals of coordination, better understanding how coordination manifests in search directives would be a promising avenue for future research. 
We note that not all of the queries associated with the CIP require a banner at the time of this writing, as some were related to keywords whose search rankings have become highly competitive over time, such as ``weaponized artificial intelligence.''

\begin{figure*}[t!]
    \centering
    \begin{subfigure}[t]{0.40\textwidth}
    \includegraphics[width=\linewidth]{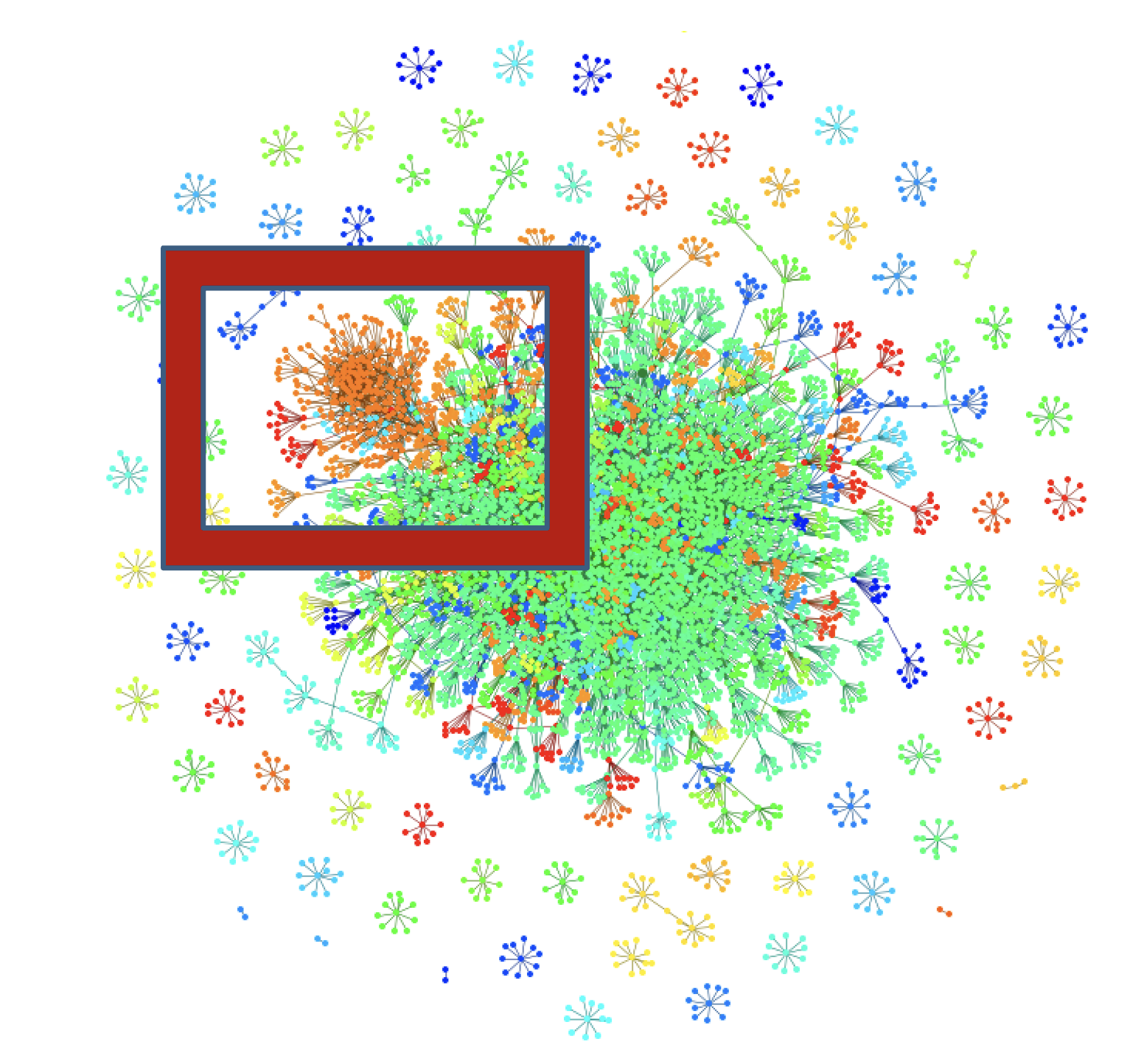}
    \caption{Graph of the domains returned in low-quality SERPs and the 10 websites that 10 domains that most frequently link to each of them.}
    \label{fig:serp_network}
    \end{subfigure}
    ~
    \begin{subfigure}[t]{0.56\textwidth}
    \includegraphics[width=\linewidth]{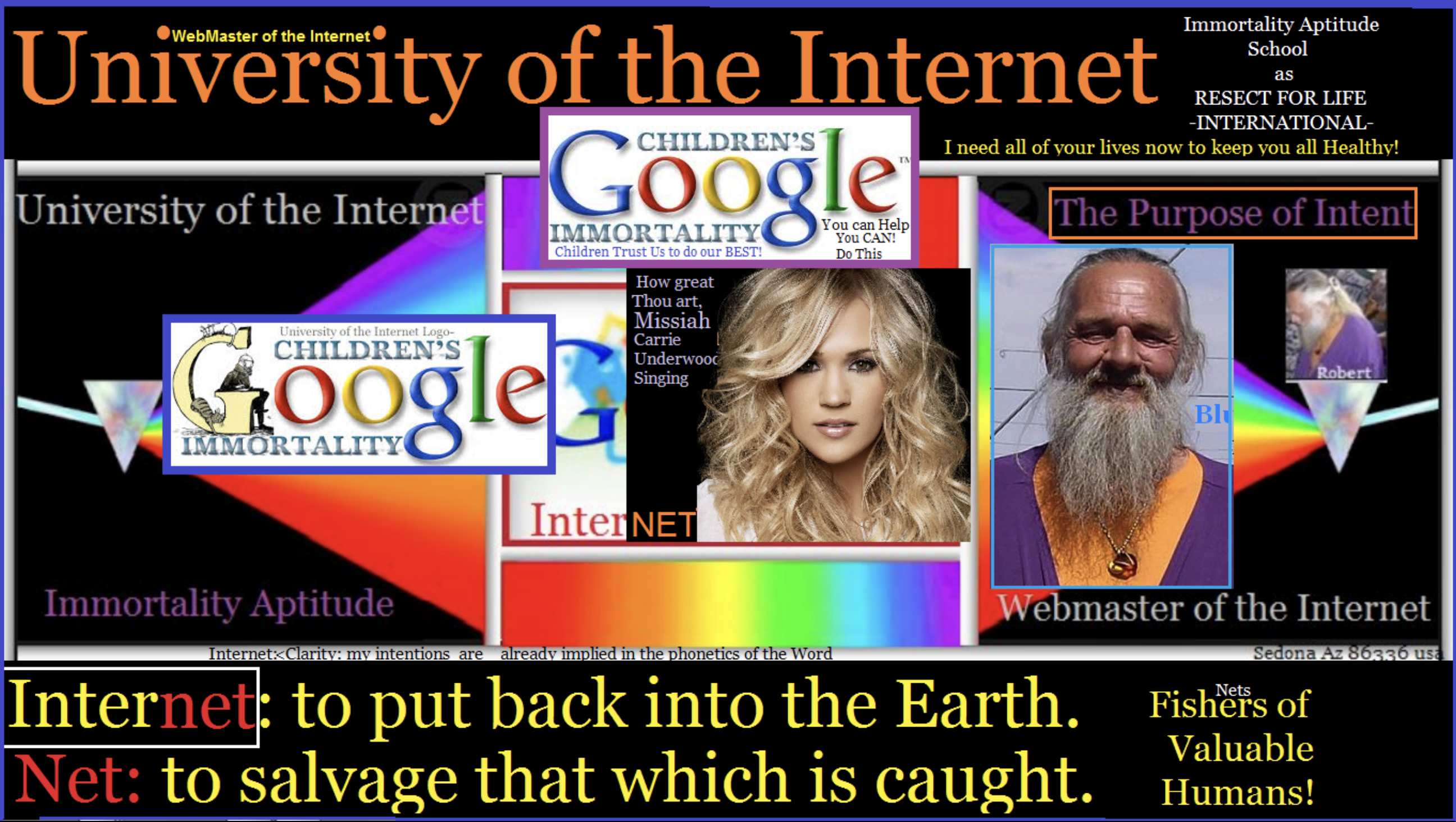}
    \caption{The old landing page of the google-bomb.com, one of the Children's Immortality Project Websites, available at: \url{https://web.archive.org/web/20141217044958/http://google-bomb.com/}}
    \label{fig:cipwebmaster}
    \end{subfigure}
    \hfill
    \caption{Children's Immortality Project SERP network and landing page.}
    \label{fig:app_cip}
\end{figure*}

\end{appendices}

\clearpage
\newpage

{\small \printbibliography}

\end{document}